\documentclass[matbio]{svjour}
\usepackage{amsmath,amssymb}
\usepackage{graphicx,epsfig,pstcol,psfrag,array}

\def\be{\begin{equation} \label}
\def\ee{\end{equation}}

\newcommand{\dd}{{\, \rm d}}
\def\l{\ell}
\def\ti{\to\infty}
\renewcommand{\sp}[1]{\langle #1\rangle}

\newcommand\iN{{\text{\small$ \frac{i}{N}$} }}
\newcommand\biN{\Big(\iN\Big)}
\newcommand\cON{\cO \Big (\text{\small $\frac{1}{N}$} \Big )}
\newcommand\cOsN{\cO \Big (\text{\small $\frac{1}{\sqrt N}$} \Big )}

\def\Poi{\text{\rm Poi}}
\def\Bin{\text{\rm Bin}}
\def\Mult{\text{\rm Mult}} 

\def\a{\alpha}
\def\b{\beta}
\def\d{\delta}
\def\eps{\varepsilon}

\def\k{\kappa}
\def\ph{\varphi}
\def\s{\sigma}
\def\t{\tau}

\def\vS{\varSigma}

\def\1{\mathsf{1}}
\def\sA{\mathsf{A}}
\def\sB{\mathsf{B}}
\def\sC{\mathsf{C}}
\def\sD{\mathsf{D}}

\def\sH{\mathsf{H}}
\def\sP{\mathsf{P}}

\def\A{\boldsymbol{A}}
\def\E{\boldsymbol{E}}
\def\F{\boldsymbol{F}}
\def\G{\boldsymbol{G}}

\def\I{\boldsymbol{I}}

\def\R{\boldsymbol{R}}
\def\U{\boldsymbol{U}}

\def\P{\boldsymbol{P}}
\def\p{\boldsymbol{p}}

\def\EE{\mathbb{E}}
\def\PP{\mathbb{P}}
\def\XX{\mathbb{X}}
\def\II{\mathbb{I}}
\def\RR{\mathbb{R}}
\def\ZZ{\mathbb{Z}}

\def\tA{\widetilde{\A}}

\def\id{\mathrm{id}}

\def\cB{\mathcal{B}}

\def\cD{\mathcal{D}}

\def\cO{\mathcal{O}}
\def\cR{\mathcal{R}}
\def\fS{\mathfrak{S}}
\def\cU{\mathcal{U}}

\def\cP{\mathcal{P}}

\def\e{{\rm e}}

\let\oldcite\cite
\def\cite{\upshape\oldcite}
\smartqed
\makeatletter
\newif\if@showqed
\@showqedtrue
\global\@namedef{endproof}{\if@showqed\qed\fi\global\@showqedtrue\@endtheorem}
\makeatother
\makeatletter
\newlength\qedraise
\newcommand\qedhere{\@ifnextchar[{\qed@here}{\qed@here[0pt]}
}
\def\qed@here[#1]{%
  \global\setlength{\qedraise}{#1}%
  {\@xp\aftergroup\csname\@currenvir @qed\endcsname}\global\@showqedfalse}%
  \def\displaymath@qed{%
    \relax
    \ifmmode
      \ifinner \aftergroup\linebox@qed
      \else
        \eqno
        \let\eqno\relax \let\leqno\relax \let\veqno\relax
        \raisebox{\qedraise}{\qed}%
      \fi
    \else
       \aftergroup\linebox@qed
    \fi
  }
  \@xp\let\csname equation*@qed\endcsname\displaymath@qed
  \@xp\let\csname multline*@qed\endcsname\displaymath@qed
  \def\align@qed{\tag*{\raisebox{\qedraise}{\qed}}}
  \@xp\let\csname align*@qed\endcsname\align@qed
\def\linebox@qed{\hfil\hbox{\qed}\hfilneg}
\makeatother

\begin{document}
\bibliographystyle{abbrv}

\title{Mutation, selection, and ancestry in branching models: 
a variational approach}

\author{Ellen Baake \inst{1} \and Hans-Otto Georgii \inst{2}}
\institute{
 Faculty of Technology, Bielefeld University, Postfach 100131,
D-33501 Bielefeld, Germany, \email{ebaake@techfak.uni-bielefeld.de}
\and
Department of Mathematics, University of Munich,
 Theresienstr. 39, D-80333 M\"{u}nchen, Germany,
\email{georgii@math.lmu.de}
}
\titlerunning{Mutation, selection and ancestry in branching models}

\makeatletter
\def\@date{}
\makeatother
\def\copyleft{}
\def\makeheadbox{{%
\hbox to0pt{\vbox{\baselineskip=10dd\hrule\hbox
to\hsize{\vrule\kern3pt\vbox{\kern3pt
\hbox{\bfseries DRAFT}
\hbox{\today}
\kern3pt}\hfil\kern3pt\vrule}\hrule}%
\hss}}}
\maketitle

\begin{abstract}
We consider the evolution of populations under the joint action of
mutation and differential reproduction, or selection. The population is
modelled as a finite-type Markov branching process in continuous time,
and the associated genealogical tree is viewed both in
the forward and the backward direction of time. The stationary
type distribution of the reversed process, the so-called ancestral 
distribution,
turns out as a key for the study of mutation-selection
balance. This balance can be expressed in the form of a variational principle
that quantifies the respective roles of reproduction and mutation
for any possible type distribution. It shows that the mean growth rate of 
the population
results from a competition for a maximal
long-term growth rate, as given by the difference between 
the current mean
reproduction rate,  and an asymptotic decay rate related to the
mutation process; this tradeoff is won by the ancestral distribution.

We then focus on the case when the type is determined by a
sequence of letters (like nucleotides or matches/mismatches
relative to a reference sequence),  and we ask how 
much of the above competition
can still be seen by observing only the letter composition
(as given by the frequencies of the various letters within the sequence).

If mutation and reproduction rates can be approximated in a smooth way,
the fitness of letter compositions resulting from the interplay of
reproduction and mutation is determined in the limit
as the number of sequence sites tends to infinity.

Our main application is the quasispecies model of sequence evolution
with mutation coupled to reproduction but independent across sites,
and a fitness  function that is invariant
under permutation of sites. In this model, the
fitness of letter compositions is worked out explicitly. In certain
cases, their competition leads to a phase transition.
\end{abstract}


\section{Introduction}

Evolution is often understood as an optimization process of some
kind, and there is a long tradition to consider evolutionary 
models, particularly those from population genetics, from a
variational perspective.  The most popular result in this context
is known as the  Fundamental Theorem of Natural Selection (FTNS). In 
its simplest form it
states that, in the deterministic selection equation for a single
locus in continuous time,
mean fitness can only increase along trajectories (i.e., it is a
Lyapunov function), and the rate of this increase equals the variance
in fitness, cf. \cite[Ch.~I.10.3]{Buer00}. More sophisticated 
versions in the context
of quantitative genetics and multiple loci, along with a
general discussion of optimality principles for the selection
equation, are discussed in \cite[Ch.~II.6.3--II.6.6]{Buer00} and
\cite[Ch.~2.9, 7.4.5 and 7.4.6]{Ewen04};
see also \cite{Edwa02}.

If, rather than selection alone, the joint dynamics of selection
and mutation is considered, results become  sparse. 
The FTNS may be generalized to house-of-cards mutation 
(i.e., mutation rates are independent of the parent type),
see \cite{Akin79} and \cite{Hofb85}.
If mutation is reversible, a Lyapunov function is available
for a certain $L^2$-renormalized version of the dynamics,
but not for the original mutation-selection equation \cite{Stan04}.

The above approaches refer to the genetic (or, more generally, type)
composition at the \emph{population level}. In contrast, this article
is concerned with a variational principle
in mutation-selection models (and closely related branching processes)
from the point of view of individual lineages
through time, their ancestry and genealogy. This principle
is  related to the (stochastic) processes
that take place along such lines of descent, with a special emphasis on the
relation between the present and the past. We will, however, not
include genetic drift (i.e., resampling) into our models;
therefore, our backward point of view  differs from that of the
coalescent process (see \cite{HSW05} for a recent review of this area). 

The paper is organized as follows. 
In Section~\ref{sec:models}, we will set up our model(s) and recapitulate
a few fundamental facts. Section~\ref{sec:preview} provides an informal
preview of the results that will be detailed (and proved) in the
remainder of the article. Section~\ref{sec:trees} will develop the lineage
aspect that will be required furtheron. 
Looking at the mutation process along
individual lines, we will obtain a fairly general
variational principle (Section \ref{sec:varprinc}, 
Thm.~\ref{thm:fundamental_eq}),
which quantifies the tradeoff between the mean reproduction rate
along a line and the asymptotic rate at which it is lost; it further implies 
a connection between the type processes that emerge in the
forward and backward directions of time. 
In Sections \ref{sec:contract} and \ref{sec:asymptotics}, we will specialize
on the case where types are sequences  over a finite
alphabet. If mutation is independent and fitness is additive across sites,
the original high-dimensional variational principle may be reduced to a
simpler, low-dimensional one (Section  \ref{subsec:linear}, 
Thm.~\ref{thm:Lambda_linear}). The same holds
asymptotically if mutation rates and fitness function
allow for a suitable smooth approximation 
when the number of sites gets large (Section \ref{sec:asymptotics}, 
Thms.~\ref{thm:lambda_asym}, \ref{thm:Lambda_e} and \ref{thm:basin}).
The corresponding approximate maximum principle will be derived
explicitly for the quasispecies model of sequence evolution (Section 
\ref{sec:quasispec}, Thm.~\ref{thm:quasispec}). 

The paper ties together, unifies and generalizes various aspects that
have appeared 
in previous publications. Special cases of the low-dimensional 
maximum principle
were first described in \cite{HRWB02}, and applied to concrete
examples. An extension  appeared in \cite{BBBK05};  it relies on  methods from 
linear algebra and asymptotic
analysis, but makes no connection to the stochastic processes on 
individual lines, nor does it include worked  examples.
The connection to  the backward point of view
relies on earlier work on branching processes
\cite{Jage89,Jage92} and
was investigated in \cite{HRWB02} and \cite{GeBa03}. These results
will reappear here as parts of a larger picture.

\section{Models and basic facts}
\label{sec:models}

\subsection{Models}

Consider a finite set of types $S$ (with $|S| > 1$) and a population 
of individuals, each of which carries one of these types.
(We think of individuals as haploid, and of types as alleles.)

\begin{figure}[h]
  \psfrag{D}{$D_i$}
  \psfrag{B}{$B_i$}
  \psfrag{i}{$i$}
  \psfrag{j}{$j$}
  \psfrag{U}{$U_{ij}$}
  \begin{center}
  \includegraphics{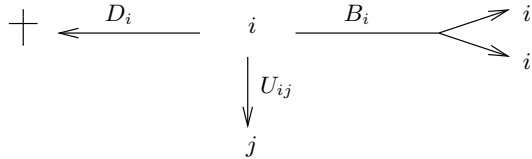}
  \end{center}
  \caption{\label{fig:museP} The parallel mutation-reproduction model.}
\end{figure}

\smallskip
\emph{\thesubsection.1 The parallel mutation-reproduction model}. 
Let us start with the most basic mutation-reproduction model in which
mutation and reproduction  
occur in parallel, that is, independently.
As depicted in Fig.~\ref{fig:museP},
an individual of type $i \in S$ may, at every instant in continuous time, 
do either of three things: It may split, i.e., produce a copy
of itself (this happens 
at birth rate $B_i \geq 0$), it may die (at rate $D_i \geq 0$), 
or it may mutate to 
type $j \; (j \neq i)$ (at rate $U_{ij} \geq 0$). Different meanings may be
associated with this verbal description. Probabilists will take it
to mean a \emph{multi-type Markov branching process in continuous time} (see
\cite[Ch.\ V.7]{AtNe72},  or \cite[Ch.\ 8]{KaTa75}
for a general overview). That is, 
an $i$-individual waits for an exponential time with parameter $A_i =B_i+D_i + 
\sum_{j:j\neq i} U_{ij}$, and then dies, splits or mutates to type $j \neq i$ with 
probabilities $B_i/A_i, D_i/A_i$, and $U_{ij}/A_i$, respectively.
The number of 
individuals of type $j$ at time~$t$, 
$Z_j(t) \in \ZZ_{\geq 0}:=\{ 0,1,2 \dots \}$, 
is a random variable; the collection  $Z(t)=\big(Z_j(t)\big)_{j \in S}$ is a random vector.
The corresponding expectation is described by the 
first-moment generator $\A=\U+\R$. Here, $\U$ is the Markov generator 
$\U=(U_{ij})_{i,j\in S}$, where the mutation rates $U_{ij}$ for $j \neq i$
are complemented by
$U_{ii}:= - \sum_{j:j\neq i} U_{ij}$ for all $i \in S$. 
Further, $\R:= \mbox{diag}\{ R_i 
\mid i \in S\}$, where $R_i:= B_i - D_i$ is the net reproduction rate 
(or Malthusian fitness). 
More precisely, we have $\mathbb{E}^i (Z_j(t)) = 
(\e^{t\A})_{ij}$, where $\mathbb{E}^i (Z_j(t))$ is the expected number 
of $j$ individuals at time $t$ in a population started by a single 
$i$-individual at time $0$.

\smallskip
\emph{\thesubsection.2 Deterministic aspects}. 
Ignoring stochastic effects and focussing on the mean behaviour of
the population, one often considers  the \emph{deterministic} 
mutation-reproduction model
\be{eq:start}
\dot{y} (t) = y(t)  \A, \qquad y(0) = y_0,
\ee
where $y(t)=(y_i(t))_{i \in S}$ is the row vector associating to each type $i$
its abundance $y_i(t) \in \RR_{\geq 0}$
(i.e., the size of the subpopulation 
of type $i$). As $y(t) = y_0 \e^{t\A}$, the deterministic 
model describes the expectation of the corresponding  branching process,
provided the initial condition is chosen accordingly.

However, the independent reproduction of individuals as implied so far
is unrealistic for large populations. They usually experience density 
regulation;
in the simplest case, this is modelled by an additional death term 
$\gamma (t) \geq 0$, that is, $D_i$ is replaced by $D_i + \gamma (t)$
(for all $i \in S$), where 
$\gamma (t)$ may depend on time (maybe through total population size), but 
not on the  type. Then, of course, \eqref{eq:start}  generalizes to
\begin{equation} \label{eq:simple}
\dot{y} (t) = y(t) (\A- \gamma (t)  \I),
\end{equation}
where $\I$ is the identity matrix. In theoretical ecology, a wide variety
of models is in use that specify  $\gamma$ for the many biological
situations that may arise.
In population genetics, however, one is usually more interested in the 
\emph{relative frequencies} $q_i (t) := y_i(t) / \sum_j y_j(t)$.
Differentiating this and inserting \eqref{eq:simple}  
leads to 
\begin{equation}\label{eq:muse}
\dot{q}_i (t) = q_i(t) \left(R_i - \sp{q(t),R} \right) + 
\sum_{\substack{j \in S:\\j\neq i}} 
\big ( q_j (t) U_{ji} - q_i (t) U_{ij} \big ),
\end{equation}
\emph{independently} of $\gamma$.
Here we think of the row vector $q=(q_i)_{i \in S}$
as a probability measure, of the 
column vector $R=(R_i)_{i \in S}$  as a function on $S$
(known as the \emph{fitness function}), and 
of the scalar product $\sp{q(t),R} = \sum_{i\in S} q_i(t) R_i$ 
as the associated expectation, namely, the mean fitness of the population
at time $t$.
Eq.~\eqref{eq:muse} is the well-known \emph{parallel} (or \emph{decoupled}) 
\emph{mutation--selection model},
which goes back to \cite[p.~265]{CrKi70}.
Although we have derived it here for haploid populations
(and will adhere to this picture),
it is well known, and easily verified, that the same equation 
describes diploids without dominance (in an approximation using 
Hardy-Weinberg proportions).
For a comprehensive review of the model and its properties, 
see \cite[Ch.~III]{Buer00}.

Rather than considering deterministic and stochastic models separately, 
we aim at a unifying picture and note that the branching process 
is particularly versatile:
Its expectation fulfills \eqref{eq:start},
and the solution of \eqref{eq:start}, in turn, implies that of
\eqref{eq:muse} (via normalization). Properties of the branching
process will, therefore, immediately translate into properties
of the mutation-selection equation (but not, necessarily, vice versa).
For this reason,
we will  consider the branching process as our primary model
throughout this paper.
Let us, therefore, return to branching populations and look at
alternatives to the parallel model.

\begin{figure}[h]
  \psfrag{D}{$D_i$}
  \psfrag{BP}{$B_i P_{ij}$}
  \psfrag{i}{$i$}
  \psfrag{j}{$j$}
  \begin{center}
  \includegraphics{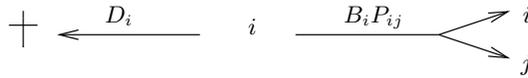}
  \end{center}
  \caption{\label{fig:museC} The coupled mutation-reproduction model.}
\end{figure}

\smallskip
\emph{\thesubsection.3 The coupled mutation-reproduction model}. 
In this model one assumes that mutations occur on the occasion of reproduction 
events (see Fig.~\ref{fig:museC}): An $i$-individual again dies at rate 
$D_i$ and gives birth at rate $B_i$, but  every time it gives birth, 
the offspring is possibly mutated (of type 
$j$ with probability $P_{ij}$), while the parent itself survives
unchanged. The corresponding first-moment generator 
$\A$ has elements 
\be{eq:A_coupled}
  A_{ij} = B_i P_{ij} - D_j.
\ee
An example of the coupled model will be studied in Sec. \ref{sec:quasispec}.

\begin{figure}[h]
  \psfrag{i}{$i$}
  \psfrag{underbrace}{$\underbrace{\mbox{\hspace{8em}}}_{\textstyle \tau_i}$}
  \psfrag{Nij}{\raisebox{.5ex}{\mbox{\large $\Bigg \}$}} $\, N_i = (N_{ij})_{j \in S}$}
  \begin{center}
  \includegraphics{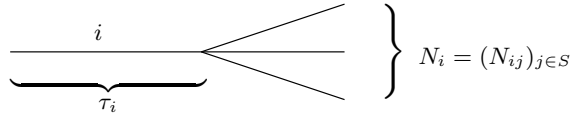}
  \end{center}
  \vspace{1em}
  \caption{\label{fig:general}General splitting rules.}
\end{figure}

\smallskip
\emph{\thesubsection.4 General splitting rules}. 
Both the parallel 
and the coupled models are special cases of the {general Markov branching 
model} as depicted in Fig.~\ref{fig:general}: An $i$-individual lives for an 
exponential time $\tau_i$ with prescribed parameter $A_i$ and then produces a 
random offspring $N_i = (N_{ij})_{j\in S}$ with distribution $\p_i$ on 
$\mathbb{Z}_+^{S}$ and finite means $\mathbb{E} (N_{ij})$ for all
$i,j \in S$. More precisely, $N_{ij} \in \ZZ_{\geq 0}$ is the 
number of children of type $j$,  and
$\p_i(\kappa) = \PP(N_{ij}=\kappa_j \,, \forall j \in S)$.
The first-moment generator $\A$ has elements $A_{ij} 
= A_i (\mathbb{E} (N_{ij}) - \delta_{ij})$.

For the coupled and the general branching rules, the first-moment 
generator may again be written in the `parallel' form $\A=\U+\R$ 
where $\U$ is a Markov generator and $\R$ is a diagonal matrix; 
this decomposition is uniquely given by $U_{ij} = A_{ij} \mbox{ for } 
i \neq j$, $U_{ii} = - \sum_{j:j\neq i} U_{ij}$, and
$ R_i = \sum_{j \in S} A_{ij}$ for all $i \in S$.
At the time being, this  is 
a formal decomposition, but will receive its branching 
process interpretation later in Sec.~\ref{subsec:sizebias}.   
The corresponding deterministic models then all take the form 
\eqref{eq:start} and \eqref{eq:muse}, provided the parameters are 
interpreted in the above way.

\subsection{Fundamental facts}

\emph{\thesubsection.1 Forward view and long-time characteristics}. 
We will assume throughout that $\A$ (or, equivalently, $\U$) 
is irreducible. Perron-Frobenius 
theory then tells us that 
$\A$ has a principal eigenvalue $\lambda$ (namely a real eigenvalue 
exceeding the real parts of all other eigenvalues) and associated 
positive left and right eigenvectors $\pi$ and $h$ which will be 
normalized so that $\langle\pi,\1\rangle = 1 = \langle \pi, h \rangle$, 
where $\1 = (1)_{i\in S}$ is the vector with all coordinates equal to $1$.
We will further  assume  that $\lambda >0$,
i.e., the branching process is \emph{supercritical}. This implies that 
the population will, in expectation,  grow in the long run, as  is obvious
from \eqref{eq:start}; in individual realizations,
it will survive with positive probability, and then grow to
infinite size with probability one, see \eqref{eq:growthrate} below. 

The asymptotic properties of our models forward in time  are, to a large 
extent, determined by $\lambda , \pi$, and $h$, and provide further connections
between the stochastic and the deterministic pictures. 
The left eigenvector $\pi$ holds the stationary 
composition of the population, in the sense that $\lim_{t\to\infty}
q(t) = \pi$ for the differential equation \eqref{eq:muse},
and, for the branching process,
\be{eq:pi}
\lim_{t\to \infty}
\frac{Z(t)}{\| Z(t)\|_1} = \pi \quad \text{with probability one,
conditionally on survival},
\ee
where $\| Z(t) \|_1  := \sum_{j \in S} Z_j(t)$ is the total population
size. 
This is due to the famous  
Kesten-Stigum theorem, see \cite{KeSt66} for the discrete-time original,
and \cite[Thm.~2, p.~206]{AtNe72}  and 
\cite[Thm.~2.1]{GeBa03} for  continuous-time versions.
Furthermore,
\be{eq:growthrate}
  \sp{\pi,R} = \lambda = \lim_{t \to \infty} \frac{1}{t} \log \| y(t) \|_1 
  = \lim_{t \to \infty} \frac{1}{t} \log  \| Z(t) \|_1
\ee
is the  asymptotic growth rate (or equilibrium mean fitness) of
the population. Here the first equality follows from the identity
$\lambda=\sp{\pi \A,\1}=\sp{\pi, \A\1}=\sp{\pi,R}$;
the second one is an immediate consequence of \eqref{eq:start} 
and Perron-Frobenius theory, and the third
is from \cite{GeBa03} and holds with probability one
in the case of survival.
Finally, the $i$-th coordinate $h_i$ of the
right eigenvector $h$ measures  the asymptotic 
mean offspring size of an $i$ individual, relative to the total size of the
population:
\be{eq:h}
h_i = \lim_{t\to \infty} \EE^i \big ( \|Z(t)\|_1 \big ) \e^{-\lambda t}.
\ee 
For more details concerning  this quantitity, see 
\cite{HRWB02} and \cite{GeBa03} (for the deterministic and stochastic
pictures, respectively).

\begin{figure}[h]
  \psfrag{t}{$t$}
  \psfrag{ttau}{$t-\tau$}
  \begin{center}
  \includegraphics{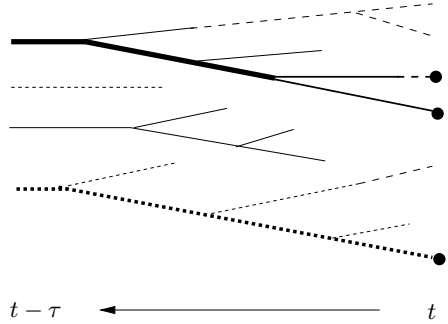}
  \end{center}
  \caption{\label{fig:backward} The backward point of view. The
  various types are indicated by different line styles.
  The fat lines mark the lines
  of descent defined by three individuals (bullets) picked from
  the branching population at time $t$. After coalescence of
  two such lines, the common ancestor receives twice the `weight', as
  indicated by the extra fat line; this motivates the factor
  $h_i$ in the ancestral distribution.}
\end{figure}

\smallskip
\emph{\thesubsection.2 Backward view and ancestral distribution}. 
In the above, we have adopted the traditional view on branching processes,
which is forward in time. It is less customary, but equally rewarding,
to look at branching populations backward in time. To this end,
consider picking individuals randomly (with equal weight) from the
current population and tracing their lines of descent backward in time
(see Fig.~\ref{fig:backward}). 
If we pick an individual at time $t$ and ask for the probability that
the type of its ancestor is $i$ at an earlier time $t-\tau$, the answer
will be $\alpha_i = \pi_i h_i$ in the limit when first $t \ti$ and then $\tau \ti$.
Thus the distribution $\alpha=(\alpha_i)_{i \in S}$ describes the
\emph{population average} of the ancestral types and is
termed the \emph{ancestral distribution}, see \cite[Thm.~3.1]{GeBa03}
for details. Likewise, the \emph{time average} along  ancestral
lines also converges to $\alpha$ in the long run, see \cite[Thm.~3.2]{GeBa03}.

If we pick individuals from the population at a very late
time (so that its composition is given by the stationary vector $\pi$),
then the type process in the backward direction is the Markov chain with
generator $\bar \G=(\bar G_{ij})_{i,j \in S}$, 
$\bar G_{ij} = \pi_j (A_{ji} - \lambda \delta_{ij}) \pi_i^{-1}$,
as first identified by Jagers \cite{Jage89,Jage92}. 
The corresponding time-reversed process has generator $\G=(G_{ij})_{i,j \in S}$, where
\be{eq:G}
G_{ij} = \alpha_j \bar G_{ji} \alpha_i^{-1} =
h_i^{-1} (A_{ij} - \lambda \delta_{ij}) h_j;
\ee
it  has been considered in
\cite{GeBa03}, has been termed
the \emph{retrospective process}, and may be 
understood as the forward type process along the ancestral lines leading
to typical individuals of the present population.
By definition, $\G$ and $\bar \G$ both have
stationary distribution $\alpha$.

\section{Preview of results}
\label{sec:preview}

In this Section, we will give an informal preview of the results that
will be obtained in the remainder of the article. This overview will
not aim at full generality, nor will it dwell on specific technical
conditions that are required to make things precise. Rather, we will
try and motivate the concepts and explain the results in the context
of the  model. The  details will be worked out in the later Sections.

We will work our way from the more general to the more specific.
We will start with a general variational principle,
valid for all model variants of the previous Section, irrespective
of the type space and of the parameters. Next, we will specialize
on the case where types are sequences, and
mutation and reproduction rates  are invariant under
permutation of sites.
This will allow to dissect  the variational
problem into two simpler problems, which are  easier to solve. 
Finally, we will treat one specific
example, namely, the quasispecies model of sequence evolution, in full detail. 

\subsection{The general variational principle}

A main object of this paper is to show that the asymptotic growth rate 
$\lambda$ of the population can be understood as the result of a competition
between the  mutation  and  reproduction processes
along a typical ancestral line. 
In this informal Section, however, we will avoid the family tree picture, and 
rather imagine we are observing just one line.
To start with, we even ignore reproduction, and consider only the
simple Markov process $\{M(t)\}_{t \geq 0}$ on $S$
with generator $\U$; i.e., the type process 
which associates with $t$ the type at time $t$ under the mutation model $\U$.
A crucial quantity in what follows will be the corresponding
\emph{empirical measure} 
\be{eq:L_preview}
  L(t):= \frac{1}{t} \int_0^t \delta_{M(\tau)} d\tau,
\ee
i.e., the random vector with components 
$L_i(t):= \frac{1}{t} \int_0^t \II\{M(\tau)=i\} d\tau$,
where $\II \{ . \}$ denotes the indicator function.
This quantity measures the fraction of 
time the  process spends in the various
states, and hence is also known as \emph{occupation time measure}. Clearly,
$L(t)$ is a random element of $\sP(S)$, the set of all probability
measures on $S$. It is well-known by the ergodic theorem for Markov chains
that, for $t \to \infty$, one has $L(t) \to \rho$ with probability
one, where $\rho$ is the
stationary distribution of $\U$. It is, perhaps, less 
well-known that the \emph{rate} of convergence may be characterized 
asymptotically
by a so-called {\em large deviation principle}, which may be informally
put as
\be{eq:LDP_preview}
  \PP(L(t) \sim \nu) \approx \e^{-t I_{\U}(\nu)} \quad
  \text{for large} \; t,
\ee
that is, the probability that $L(t)$ is close to some measure $\nu$
decays exponentially, for large time, with a decay rate (or \emph{rate function})
$I_{\U}(\nu)$  which can be written down explicitly (see \eqref{eq:I_U} and 
\eqref{eq:Iroot} below).
$I_{\U}$ is  nonnegative, and
$I_{\U} (\nu) = 0$ precisely for $\nu=\rho$, in line with the above fact that,
in the long run, only the stationary measure $\rho$ will survive.

Let us now add  reproduction, i.e., turn to the branching process. 
As a consequence of the above large deviation principle, we will obtain, in 
Thm.~\ref{thm:fundamental_eq}, a link between
the forward-time stationary 
distribution $\pi$ of the branching process,  the reproduction rate 
$R$, the asymptotic growth rate~$\lambda$, the mutation process
$\U$ and the ancestral distribution $\alpha$, namely, the equation
\be{eq:varprinc_alpha_preview}
 \langle \pi, R \rangle = \lambda 
  = \max_{\nu \in \sP(S)}\big [\langle \nu,R \rangle - I_{\U}(\nu)\big ]
          = \langle \alpha, R \rangle - I_{\U}(\alpha)\,.
\ee
This variational principle may be understood in terms of a
competition between all possible distributions for a maximal
long-term growth rate, as given by the difference between 
the current mean
reproduction rate $\langle \nu, R \rangle$, and the asymptotic decay rate 
$I_{\U}(\nu)$.
The first quantity is maximized by those measures that
put mass only on the fittest type(s); the second one is minimized
by $\rho$; the tradeoff is won by $\alpha$. 
Furthermore, \eqref{eq:varprinc_alpha_preview}
connects the forward and the retrospective point of view in that
the maximum equals the mean fitness $\langle \pi,R \rangle$ of the
stationary population. Note that the mean fitness of the ancestral population
exceeds the mean fitness of the stationary one by $I_{\U}(\alpha)$, which
is positive unless $\alpha=\rho$ (which implies
$R_i= \text{const.}$, i.e., there is no selection).
This reflects the fact that the present population carries with it a tail of 
(mainly unfavourable) mutants that are present at any time, but do not 
survive in the long run.

We will see in Sec.~\ref{sec:varprinc} that this 
`competition of distributions' 
can be made more concrete, namely, in terms of a competition of 
lines of descent,  by considering the empirical distributions $L^\omega(t)$
of types along distinct lines $\omega$. But before we can embark on this, we must first 
develop a way of constructing trees, lines, and processes on lines,
in a consistent way; this will be taken up in the next Section.

It is interesting to note that the above variational principle resembles
the thermodynamic maximum principles
in statistical physics. Indeed, our reproduction rates may be identified with an energy,
and the rate function with an entropy; in fact, the rate function
for the continuous-time Markov chain $M(t)$ can be naturally derived
from the usual entropy governing the
so-called pair-empirical measure of a discrete-time Markov chain, cf.\ 
\cite[Ch.~IV]{deHo00}.

\subsection{Sequence space models}
\label{sec: SequenceSpace}
The variational principle \eqref{eq:varprinc_alpha_preview}, valuable as
it is conceptually, is not very useful if one aims at 
an explicit
solution; this is because maximization is over a large space
(the set of probability measures on $S$). However, it turns out that,
in certain models of sequence evolution, this task boils down to a much
simpler one if the original problem is dissected into two, one of which
can be solved explicitly. Let us first describe this `divide
and conquer' strategy.

Assume that the type of an individual is characterized by a  sequence
of nucleotides, amino acids, matches/mis\-matches with respect to
a reference sequence of nucleotides, or, in general,  letters from some 
alphabet $\varSigma$. Thus $\varSigma = \{A,G,C,T\}$,  $ \{1,\ldots,20\}$, 
$ \{0,1\}$, or  any other
finite set\footnote{As in the case of matches/mis\-matches, the 
formal alphabet $\varSigma$ need
not coincide with the alphabet used in the biological description.
The letters in the original biological sequence may, for example,
even be replaced by 
$n$-tuples of matches/mismatches relative to $n$
reference sequences, as required in the treatment of Hopfield
fitness functions \cite{BBBK05,Gars05}}. 
The natural type space is then $\varSigma^N$, the set of  possible
sequences of length $N$, where $N$ is typically large. However, if the
mutation and reproduction rates are invariant under permutations of  sequence sites,
all relevant information on a sequence $\s=(\s_k)_{1\le k\le N}\in \varSigma^N $ is
already contained in its \emph{letter histogram} 
(or \emph{letter composition}) 
\begin{equation}\label{eq:H}
H(\s)= (H_\l(\s))_{\l \in \varSigma}\,, \quad H_\l(\s)=\sum_{k=1}^N \II\{\s_k=\l\},
\end{equation}
which indicates how often each letter~$\l$ shows up in $\s$.
In other words, it is sufficient to look at the reduced type space
\begin{equation}\label{eq:simplex_S_preview}
S=H(\varSigma^N)=\Big\{ i \in \ZZ^{d} \mid i_{\ell} \geq 0 \: \text {for} \: \ell \in \varSigma, \;
\sum_{\ell \in \varSigma} i_{\ell} = N\Big \},
\end{equation}
with $d=\lvert \varSigma \rvert$, which consists of all possible letter
compositions. 
\begin{figure}[h]
\[
\text{\normalsize$\varSigma^N$}
\stackrel{\text{\small$H$}}{\text{\Large$\longrightarrow$}}  
\text{\normalsize$S$}
\]
\caption{\label{fig:lumping}
Lumping a sequence space.}\vspace{-1.5ex}
\end{figure}
This \emph{lumping procedure} induces a model on $S$ that is again a 
Markov branching process; its
reproduction rates $R=(R_i)_{i \in S}$ and  mutation generator 
$\U=(U_{ij})_{i,j\in S}$ are uniquely determined by the corresponding
rates of the original process on $\varSigma^N$. Many models of
sequence evolution allow for such a lumped representation; as a
particularly realistic example, let us mention the mutation-selection
model for regulatory DNA motifs \cite{GeHw02}, which also involves 
analysis of sequence data.

To get back to the variational problem,
we will classify the possible  distributions
$\nu\in\sP(S)$ according to the value of their mean $\sp{\nu,\id} \in  \RR^d$;
here $\id$ denotes the identity function on $S$ defined by $\id_i = i $ for all
$i \in S$, and, in line with previous usage,
the scalar product  gives the expectation of this
vector-valued function under the measure $\nu$.  Keeping in mind that $S$
arises from lumping a sequence space $\varSigma^N$ as in Fig.~\ref{fig:lumping}, 
we think of $\sp{\nu,\id}$ as the expected value of a random letter composition
with distribution $\nu$,
i.e., the mean histogram if histograms have distribution $\nu$.   

Let us now foliate the variational problem
\eqref{eq:varprinc_alpha_preview} according to these mean
letter frequencies. That is, we write
\be{eq:lambda_Lambda_nonstrict_preview}
   \lambda = \max_{z \in \text{conv}\: S}  \Lambda(z),  
\ee
with $\Lambda: \,  \text{conv}\: S \to \RR$ given by
\be{eq:Lambda_preview}
   \Lambda(z) := 
   \max_{\substack \nu \in \sP(S): \\ \langle \nu,\id \rangle = z} \;
   [\sp{\nu,R} - I_{\U}(\nu)].
\ee
Here we write 
$\text{conv}\: S$ for the  convex hull of $S$, that is, the set
of convex combinations of elements of $S$, or, in other words,
the set of all possible mean letter compositions. 
The (unique) maximizer of $\Lambda$ is
$\hat z := \langle \alpha, \id \rangle$,
i.e., the mean ancestral letter composition. 

The function $\Lambda(z)$ describes the growth rate resulting from
the competition between
all distributions with mean letter composition $z$; we will therefore
call it the \emph{constrained mean fitness} of $z$.
In analogy with the interpretation of the unconstrained variational principle 
\eqref{eq:varprinc_alpha_preview}, the competing distributions
may be identified with empirical  letter compositions along lines of descent,
and $\Lambda(z)$ will turn out as the \emph{asymptotic growth rate of the lines with 
empirical letter histogram (close to) $z$}; this will be shown in 
Prop.~\ref{prop:constrained_growth_rate}.
It follows that the growth rate of the total
population coincides with the growth rate $\Lambda(\hat z)$ of the 
subpopulation consisting of all lines with empirical letter histogram close 
to the mean ancestral one.

Now, the main point is that $\Lambda(z)$ can be calculated explicitly 
in two interesting situations, namely:

\medskip\noindent
(1) \emph{All sites of the sequence mutate independently and according to the
same Markov process in continuous time, and  
fitness is additive across sites}. Thus
$R_i=R(i)$ and $U_{ij}=U_{j-i}(i)$
are linear functions on $S$ (that will be extended to $\text{conv}\: S$). In Thm.~\ref{thm:Lambda_linear}, we then obtain 
$\Lambda(z)$ explicitly, and exactly,
as
\be{eq:Lambda_linear_preview}
  \Lambda(z)   = R(z) - \frac12\sum_{k}
  \Big ( \sqrt{U_k(z)} - \sqrt{U_{-k}(z)} \Big ) ^2
    = \langle \nu^{(z)}, R \rangle - I_{U}(\nu^{(z)}),
\ee
where the sum is over all possible mutational steps, and
$\nu^{(z)} =\Mult_{N,z/N}$ is the multinomial distribution with  mean $z$.

\medskip\noindent
(2) \emph{The reproduction
and mutation rates have a continuous approximation of the form
\be{eq:cont_approx_RU_preview}
R_i = r \biN + \cON \quad\text{ and }\quad U_{ij} = u_{j-i} \biN + \cON
\ee
with  functions $r$ and $u_k$ that are smooth enough}. Under further
technical conditions, an analogue of \eqref{eq:Lambda_linear_preview}
will be obtained in Thm.~\ref{thm:Lambda_e}, namely,
\begin{equation}\label{eq:Lambda_e}
   \Lambda(z) = e(z)+\cO(N^{-1/3}),
\end{equation}
where
\be{eq:e_parallel_preview}
   e(z) =  r(z) - \frac12\sum_{k} 
  \big ( \sqrt{u_k(z)} - \sqrt{u_{-k}(z)}\, \big ) ^2.
\ee
Strictly speaking, the approximation \eqref{eq:Lambda_e} is only true 
when $e(z)$ is concave; otherwise
$e(z)$ has to be replaced by its concave envelope, and the distribution
attaining the constrained maximum $\Lambda(z)$ will show distinct peaks.
This behaviour, which indicates some kind of phase transition, will be the subject 
of Thm.~\ref{thm:basin}. For $z=\hat z$, this phenomenon means that the total growth rate $\lambda$ is determined by two or more coexisting subpopulations with distinct empirical letter histograms.

\subsection{The quasispecies model} 
We will finally consider the coupled sequence space model on
$\{0,1\}^N$,  known as the 
\emph{quasispecies model}; more precisely, we will use
a slightly adapted version
of the original in \cite{Eige71}. It will be assumed that births and
deaths occur at rates that
are invariant under permutation of sites, and 
mutations occur on the occasion of birth events, independent
across sites, and at probabilities $v=\mu/N$ and $w=\nu/N$ from 
$0$ to $1$ and vice versa, where $\mu$ and $\nu$
are positive and independent of $N$. 
Then, lumping may be performed into
$S:= \{0,1,\ldots,N\}$ by counting the number of $1$'s in a sequence.
If the birth and death rates of the resulting
model on $S$ have a continuous approximation analogous to that of
\eqref{eq:cont_approx_RU_preview}, namely,
\[
B_i=b \biN+\cON \quad\text{ and }\quad D_i=d\biN+\cON,
\]
then
\be{eq:e_quasispec_preview}
e(z) := b(z) \,\exp  
       \big [- \big( \sqrt{\mu (1-z)} - \sqrt{\nu z}  \big)^2 \big ] - d(z)
\ee
takes the role of $e(z)$ in \eqref{eq:e_parallel_preview}.

\section{Trees, lines, and processes on lines}
\label{sec:trees}

To understand the probabilistic significance of the variational principle
previewed above, it is necessary to develop a detailed picture of the
branching process that includes the full family tree. However, to keep
technicalities at a minimum we confine ourselves, in the first subsection, 
to the parallel model; in this case, a
particularly simple construction is available which is sufficient
for our needs. A more versatile procedure 
for general splitting rules will be sketched
in Subsection \ref{subsec:sizebias}.

\subsection{The parallel model}
\label{subsec:parallel_tree}
\begin{figure}[h]
\begin{center}
\psfrag{A}{\small$(0,0)$}
\psfrag{B}{\small$(0,1,0)$}
\psfrag{C}{\raisebox{1.5pt}{\small$(0,1,1)$}}
\psfrag{D}{\raisebox{-2pt}{\small$(1,0)$}}
\psfrag{E}{\small$(1,1,0)$}
\psfrag{F}{\small$(1,1,1)$}
\psfrag{0}{\small$0$}
\psfrag{t}{\small$t$}
\epsfig{file=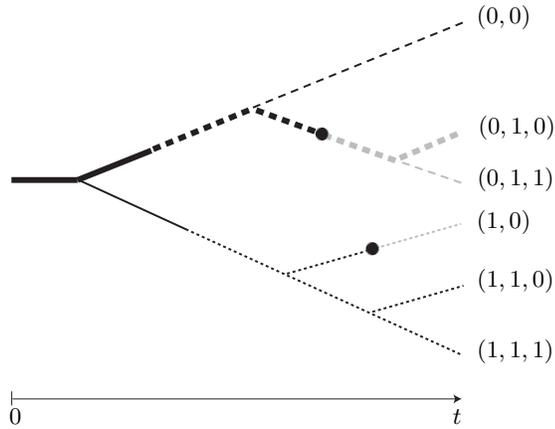, width=6.5cm}\hskip3mm
\end{center}
\caption{\label{fig:tree}
The branching process with mutation and binary
splitting. Bullets mark death events; line segments that are alive are
shown in black, virtual ones in grey, types are indicated by various  line
styles. The (randomly chosen) representative line
is marked fat; its initial segment shown here is the first child
of the second child
of the first child of the root, i.e., the individual
$x=(0,1,0)$. Since it has
experienced three splitting events, it is a third generation individual;
but it is virtual, as in fact already its mother $(0,1)$ died.
The `black' tree is a realization of the branching process.
The individuals  alive at time $t$ constitute
the population $X(t)$ (a mixture of various generations);
here, $X(t) = \{ (0,0),(1,1,0),(1,1,1) \}$; the other individuals
at time $t$ are virtual.}
\end{figure}

Let us explain the  construction for the parallel model,
as illustrated in Fig.~\ref{fig:tree}. The population is started 
by a single individual (the \emph{root})
of type~$i$. In a first step, we ignore 
all death events and consider only the splitting events. Then all lines are
\emph{infinite} and can be labeled by a  sequence $\omega \in \{0,1\}^{\ZZ_{\geq 1}} =: \Omega$, where $\omega_n$ tells us
whether the $n$-th offspring corresponds to the upper ($0$) or lower ($1$)
branch in the
graphical representation of the tree, or, equivalently, whether it is
counted as `first' or `second' at  birth.
Next, \emph{individuals} are  defined as \emph{(finite) initial segments}
of the infinite lines, i.e., $x=(\omega_1, \ldots, \omega_n)$ is
an $n$-th generation individual. The empty initial string $\emptyset$ of length $0$ 
corresponds to the root and is counted as generation
$0$. 
The set $\XX := \{ \emptyset \} \cup ( \bigcup_{n \geq 1} \{0,1\}^n )$ then comprises all
individuals that may possibly occur
(and do occur as long as death events are ignored). 

A realization of the Markov branching process described 
informally in Sec.~\ref{sec:models}
may then be specified by associating with every line 
$\omega$ the times at which it splits, its type 
(as a function of time),
and the time it dies (by a death event). For convenience, the construction proceeds
in two steps: we first grow a tree
by splitting and mutation alone (with the appropriate exponential waiting times); the death events are then
superimposed in a second step to determine which lines are still
alive. This way, lines that have already died live on virtually
and may continue to divide and mutate. However, this does not influence
the lines that are alive; only these constitute the realization of the
branching process. In particular, we denote by $X(t) \in \XX$ the
set of individuals alive at time $t$; note that this is a
mixture of various generations. (We remain a bit informal here;
for one of the various possible ways of a rigorous construction, see
\cite{GeBa03}.)

For each line $\omega$, we
consider now the following families of random variables: 
$\{M^{\omega}(t)\}_{t\geq 0}$, the type
process, which associates with $t$ the type of $\omega$ at time $t$;
$\{\beta^{\omega}(t)\}_{t\geq 0}$, the number of birth events 
along $\omega$ before $t$; 
and $T^{\omega}$,  the time line $\omega$ dies (if $\omega$ survives forever,
this time is infinite).
Both the birth and the death process depend on the type process,
but not vice versa.
The crucial information on $\{M^{\omega}(t)\}_{t\geq 0}$ is contained in its
\emph{empirical measure}
\be{eq:L}
  L^{\omega}(t):= \frac{1}{t} \int_0^t \delta_{M^{\omega}(\tau)}\, d\tau\,,
\ee
cf. \eqref{eq:L_preview}.
For an individual~$x$ at time~$t$,
the empirical measure only depends on the 
initial segment of $\omega$ that describes  $x$. With this in mind, we will
sometimes also write $L^x(t)$ rather than $L^{\omega}(t)$. 

The above families of random variables are
not independent between lines (they are dependent through common
ancestry), but, by symmetry between the two offspring at every splitting
event, they share the same \emph{marginal} laws for all $\omega \in \Omega$. 
In particular, 
since mutation is not influenced by the reproduction events,
the type process on any \emph{given} line (regardless
of the others) is a copy of the mutation process generated by $\U$.
Let us choose one particular such line $\omega^*$, for example, by setting
$\omega^*=(000\ldots)$, or by tossing a coin. The line $\omega^*$ may or may not survive, but it will always be present at least virtually.
We will call
it the \emph{representative line} for reasons to become clear in a moment,
and set $\beta(t) := \beta^{\omega^*}(t)$, $M(t) := M^{\omega^*}(t)$,  
$T(t) := T^{\omega^*}(t)$, and $L(t) := L^{\omega^*}(t)$.
We will now see that, once we know the laws of these quantities, they
can tell us a lot about the entire tree.

The basic observation is that, in generation $n$, there are $2^n$ 
possible (real or virtual)
individuals, all with the same marginal laws for the random variables
just discussed. This allows us to express the expected 
population size of
a population started by a single $i$ individual at time $0$ as follows:
\be{eq:expect_popsize}
\begin{split}
 \EE^i \big (\| Z(t) \|_1 \big ) &= \EE^i \big ( \lvert X(t) \rvert \big ) =
 \sum_{n \geq 0} 2^n\, \EE^i(\II \{\beta(t)=n,T>t\}) \\
 &=
 \EE^i\big(2^{\beta(t)}\II\{T>t\}\big)\,.
\end{split}
\ee
Now, conditionally on $L(t)$, the random variables  $\{T>t\}$ and
$\beta(t)$ are independent, having probability $\exp(-t\sp{L(t),D})$
resp.\ the Poisson distribution $\Poi_{t\sp{L(t),B}}$ with parameter 
$t\sp{L(t),B}$. Therefore, 
\[
\begin{split}
\EE(2^{\beta(t)} \mid L(t)) &= \exp(t\sp{L(t),B}), \text{ and } \\
\EE(\II \{ T>t\} \mid L(t)) &= \exp(-t\sp{L(t),D})
\end{split}
\]
(both independently of the type of the root),
where the former relies on the fact that,
for a random variable $Y$ with distribution $\Poi_{\lambda}$, one 
has $\EE(2^Y)=\e^{\lambda}$.
Therefore, \eqref{eq:expect_popsize} turns into
\be{eq:lambda_L}
\begin{split}
  \EE^i\big(\| Z(t) \|_1 \big ) & = 
  \EE^i \Big ( \EE \big ( 2^{\beta(t)} \II\{T>t\} \mid L(t) \big ) \Big ) \\
  & =  \EE^i \Big ( \EE \big ( 2^{\beta(t)} \mid L(t) \big ) 
              \,   \EE \big (\II \{T>t\} \mid L(t) \big ) \Big ) \\
  & =  \EE^i ( \e^{t \sp{L(t),B}} \e^{-t\sp{L(t),D}} ) 
    = \EE^i(\e^{t \sp{L(t),R}}).
\end{split}
\ee
Note that the remaining expectation (and the outer one where 
expectations are nested) is  with respect to $L(t)$. 
We also remark that the underlying tree construction 
lurks behind the
above derivation, but in the simple case at hand
it need not be made  more  explicit. 

\subsection{General splitting rules}
\label{subsec:sizebias}
We have, so far, restricted ourselves to the decoupled model with
parallel mutation, reproduction and death. The crucial simplifiying 
feature here is the fact that, forward in time on every line,
we have a copy of the mutation process generated by $\U$.
Therefore, we could consider any line as representative.

Outside the parallel model, the decomposition $\A=\U+\R$ is formal to
start with, and the generator $\U$ has no immediate interpretation.
But with the help of a more advanced tree construction,
one can again obtain a representative line with its type process $M(t)$
generated by $\U$. 
We will only give  a rough sketch here; for the full picture
we refer the reader to \cite{GeBa03}.

\begin{figure}[h]
\psfrag{tau}{$\underbrace{\qquad\qquad\;}_{\tau_j}$}
\psfrag{pj}{$\Biggr\} N_j\sim {\boldsymbol{p}}_j$}
\psfrag{Ni}{$\Biggr\} \widetilde{N}_i \sim \widetilde{{\boldsymbol{p}}}_i$}
\psfrag{taui}{$\underbrace{\qquad\qquad\quad\qquad\;\;}_{\widetilde{\tau}_i}$}
\psfrag{j}{$j$}
\psfrag{i}{$i$}
\begin{center}
\includegraphics{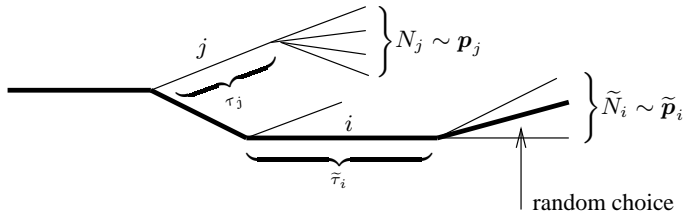}
\end{center}
\caption{\label{fig:sizebias} 
A realization of a size-biased tree with 
its trunk (the fat line). An individual of type $j$, off the trunk, has
offspring $N_j$ with distribution $\p_j$ after 
an exponential waiting time $\tau_j$  
with mean $1/A_j$; an individual
of type $i$ along the trunk bears offspring 
$\widetilde N_i$ with biased distribution  $\tilde \p_i$
after an exponential waiting time $\tilde \tau_i$ with 
mean $1/A_i \EE(\| N_i \|_1)$.}
\end{figure}

The construction relies on a so-called
\emph{size-biased tree with random spine} (or \emph{trunk}). 
The general concept was introduced in 
\cite{JaNe84,LPP95} and \cite{KLPP97}; the particular (continuous-time)
version required here can be found in
\cite[Remark 4.2]{GeBa03}. Informally, one constructs a modified
tree with a randomly selected, distinguished line (called the
trunk or spine), along which
time runs at a different rate and offspring are weighted according to
their size; in particular, there is always at least one offspring
along the trunk so that the trunk survives forever. The children off the
trunk get ordinary (unbiased) descendant trees; see Fig.~\ref{fig:sizebias}.


More precisely, for each type $i \in S$, we introduce the size-biased
offspring distribution 
\[
  \tilde \p_i(\k) = \frac{\| \k \|_1 \p_i(\k) }{\EE(\| N_i \|_1)}, 
  \quad \k \in \ZZ_{\geq 0}^S.
\] 
Starting at the root, an individual of
type $i$ on the trunk waits for an exponential time with parameter 
$A_i \EE(\| N_i \|_1)$ and then produces  offspring $\tilde N_i$
according to
$\tilde \p_i$; one of these offspring is chosen randomly (with equal weight)
as the successor on the trunk. It is easily verified that the
type process on the trunk is a Markov chain generated by $\U$.
The trunk takes the role of the representative line, and the
considerations of the previous Subsection carry over. We do not
spell this out here explicitly;  
for the complete picture and
many details, in particular on how the trunk may be used
to extract further information about the tree, see \cite{GeBa03}.
To avoid
misunderstandings, we would like to emphasize that
the size-biased tree  as applied to the {\em parallel} model 
does \emph{not} reduce to the simple special construction of the
previous Subsection.
In particular, unlike
the representative line of this construction, the trunk of the
size-biased tree is certain to survive forever. However,
both constructions share the essential property that the 
mutation process along the trunk or
representative line, respectively, is generated by $\U$,
and the fact that many properties of
the entire tree may be extracted from this distinguished line.

\section{Variational characterization of the asymptotic growth rate}
\label{sec:varprinc}

We are now in a position to derive the variational characterization
\eqref{eq:varprinc_alpha_preview} of the asymptotic growth rate $\lambda$. 
The idea is to  observe both
the mutation process and the reproduction rate along the representative
line of the tree. The appropriate tool for analyzing the tradeoff between 
these processes is the large deviation principle for the mutation process.

\subsection{Using the large deviation principle}
 
Let us, for the moment, restrict ourselves to the parallel model; 
we will see later
that our results hold automatically for general splitting rules.
For the parallel model, we can combine
\eqref{eq:growthrate} and \eqref{eq:lambda_L} to obtain
\be{eq:lambda}
  \lambda 
  =\lim_{t\to \infty}\frac{1}{t} \log \EE^i 
    (\e^{t \langle L(t), R \rangle}  )
   =\lim_{t\to \infty}\frac{1}{t} \log \EE^i 
    \Big(\exp\big[\textstyle\int_{0}^t R_{M(\tau)}\,d\tau\big]\Big)
    \,, 
\ee
that is, the growth rate can be determined by observing the types
and the associated reproduction rates along the representative line.
The competition between reproduction and mutation will lead to a variational
formula for $\lambda$, which can immediately be derived from the variational
formulas of large deviation theory. The basic fact is the following
\emph{large deviation principle} for $L(t)$, 
see 
\cite[Ch.~III.1 and IV.4]{deHo00} or \cite[Ch.~1.2 and 3.1]{DeZe98}).

\begin{proposition}\label{prop:LDP}
The empirical measure $L(t)$ of a continuous-time Markov \linebreak chain on a finite state 
space $S$ with irreducible generator $\U$ satisfies the
\emph{large deviation principle (LDP)} with rate function 
\be{eq:I_U}
  I_{\U}(\nu) := \sup_{v > 0} 
  \Big [- \Big \langle \nu, \frac{\U v}{v} \Big \rangle \Big ], \qquad\nu\in\sP(S),
\ee
where the supremum is taken over all 
$v \in \RR_{> 0}^S$,  and
the fraction is to be understood component-wise, i.e.,
$\U v/v$ is the vector with components $(\U v)_i / v_i$. 
More explicitly, the LDP means that
\[
  \limsup_{t \ti} \frac{1}{t} \log \PP \big ( L(t) \in C \big )
  \leq - \inf_{\nu \in C} I_{\U}(\nu)\
\]
for any closed set $C \subset \sP(S)$, and
\[
  \liminf_{t \ti} \frac{1}{t} \log \PP \big ( L(t) \in O \big ) 
  \geq - \inf_{\nu \in O} I_{\U}(\nu)\
\]
for any open set $O \subset \sP(S)$.
Furthermore, $I_{\U}$ is continuous, strictly convex and nonnegative, and
$I_{\U} (\nu) = 0$ precisely for $\nu=\rho$, the
stationary distribution of~$\U$. 
\end{proposition}
For an informal statement of the LDP recall \eqref{eq:LDP_preview}.
(Although we have stated the LDP here only for the special
case we need, it is indeed quite a general principle that applies 
to many common types of random variables. We refer the interested reader
to the monographs  \cite{DeZe98} or \cite{deHo00}.) 

Returning to \eqref{eq:lambda}, we now see that, on
the right-hand side, the exponential factor 
$ \e^{t \langle L(t), R \rangle}$ is  integrated over a probability measure 
that behaves essentially like $ \e^{-t I_{\U}}$. 
It may thus be evaluated by Varadhan's lemma
on the asymptotics of exponential integrals, which is a far-reaching
generalization of Laplace's method; see \cite[Thm.~III.13]{deHo00} 
or \cite[Thm.~4.3.1]{DeZe98}. Specifically, we obtain the key formula
\be{eq:varprinc}
  \lambda = \lim_{t \ti} \frac{1}{t} \log 
 \int_{\sP(S)} \e^{t \langle \nu,R \rangle} \, \PP^{i}(L(t)\in\dd\nu)
  = \max_{\nu \in \sP(S)} \big [ \langle \nu,R \rangle - I_{\U}(\nu) \big ],
\ee
which may be understood as a `largest exponent wins' principle.
Let us continue with a series of comments.

\emph{\thesubsection.1 Relation to the retrospective process}.
The maximum principle \eqref{eq:varprinc}, though derived by considering
the branching process forward in time, is directly 
connected to the retrospective process of \eqref{eq:G}. In analogy with \eqref{eq:I_U},
the rate function for the empirical measure of the retrospective process 
(generated by $\G$ of \eqref{eq:G}) reads
$I_{\G}(\nu)  = \sup_{w > 0} [- \langle \nu, (\G w)/ w  \rangle ]$. This, however,
is closely related to  $ I_{\U}(\nu)$. Indeed, setting  $v=(v_i)_{i \in S}$ with 
$v_i = h_i w_i$ we can write
$ \langle \nu, (\G w)/ w  \rangle 
   = \sum_{i,j \in S} \nu_i (A_{ij} - \lambda \delta_{ij}) h_j w_j/h_i w_i 
   = \sum_{i,j \in S} \nu_i (A_{ij} - \lambda \delta_{ij}) v_j /v_i
   = \langle \nu,R \rangle - \lambda 
     +  \langle \nu, (\U v)/ v  \rangle$, whence
\be{eq:I_G}
\begin{split}
 I_{\G}(\nu)  & = \sup_{w > 0} 
               [-  \langle \nu, (\G w)/ w  \rangle ]
             = \lambda - \langle \nu, R \rangle 
               + \sup_{v > 0} 
               [-  \langle \nu, (\U v)/ v  \rangle ] \\
             & = \lambda - \langle \nu, R \rangle  + I_{\U}(\nu) \,.
\end{split} 
\ee
Again,
$I_{\G}(\nu)$ is nonnegative, strictly convex, and vanishes if and only
if $\nu=\alpha$, the stationary distribution of $\G$.
It follows that the ancestral distribution $\alpha$ is the unique maximizer 
in \eqref{eq:varprinc}.
We may thus summarize our findings in the
following theorem (recall \eqref{eq:growthrate} for the first identity).
\begin{theorem}\label{thm:fundamental_eq}The forward-time stationary 
distribution $\pi$, the reproduction rate $R$, the asymptotic growth rate $\lambda$, 
the mutation process
$\U$ and the ancestral distribution $\alpha$ are linked via the equation
\be{eq:varprinc_alpha}
 \langle \pi, R \rangle = \lambda 
  = \max_{\nu \in \sP(S)}\big [\langle \nu,R \rangle - I_{\U}(\nu)\big ]
          = \langle \alpha, R \rangle - I_{\U}(\alpha)\,.
\ee
\end{theorem}

\emph{\thesubsection.2 The mutation rate function at the ancestral distribution}.
Thm.~\ref{thm:fundamental_eq} yields the additional relation
\[
I_{\U}(\a) =  \sp{\a,R}-\lambda  =\sp{\pi,Rh-\A h}= \sp{\pi,-\U h} =
\sum_{\substack{i,j \in S: \\i \neq j}} \pi_i U_{ij} (h_i - h_j), 
\]
i.e., the
value of the mutational rate function at the optimum equals the long-term
loss of offspring due to mutation, wherefore it was previously
termed \emph{mutational loss function};
see \cite[Sec.~5 and Appendix A]{HRWB02} for the biological
implications.

\emph{\thesubsection.3 Balance of mutation and reproduction}.
On every line $\omega$, the mutation process runs randomly through a sequence of histories, and hence determines
an evolution of empirical measures $L^\omega(t)\in\sP(S)$. As $t\to\infty$,
the empirical measures $\nu=L^\omega(t)$ that differ from the stationary
distribution $\rho$ of $\U$ become exponentially less probable at
asymptotic rate $I_{\U}(\nu)$. In particular, $\rho$ is the (almost-sure)
long-term \emph{time average}
on the line $\omega$ in the forward direction of time. 
In spite of this, the long-term 
\emph{population average} $\pi$ of \eqref{eq:pi} differs from $\rho$, in general.
This is because mutation is counterbalanced
by reproduction, at rate $R_{M^\omega(t)}$ at instant $t$, and at mean
rate $\langle L^\omega(t), R \rangle$ for the entire line segment up to time $t$.
We note that in realistic biological models the  largest
reproduction rates  typically belong to types that are  
improbable under the stationary  
mutation distribution $\rho$ (`good' types are rare under mutation alone,
otherwise it would not require selection
to establish them!).  Hence, empirical measures with a large
mean reproduction rate tend to differ markedly
from $\rho$.   The resulting  tradeoff 
between the mean reproduction rate of a line
and its asymptotic rate of decay is won by those lines $\omega$ 
for which $L^\omega(t)=\nu$ maximizes the difference, 
$\langle \nu, R \rangle - I_{\U}(\nu)$. According to 
Thm.~\ref{thm:fundamental_eq}, 
these are precisely the lines having the
\emph{ancestral distribution} $\alpha$ as their time average.
It is therefore this $\alpha$ that is successful
in the long run and that we see when looking back into the past. 

\emph{\thesubsection.4 Extension to general splitting rules}.
In our proof of Thm.~\ref{thm:fundamental_eq} above,  we used a probabilistic argument
that relied on the parallel model and the associated tree construction.
So it might seem that this theorem is limited to this particular model.
Note, however, that all quantities appearing in Thm.~\ref{thm:fundamental_eq} are
solely determined by the first-moment generator $\A$ of the process, so that it is
a property of $\A$ rather than the underlying process.  
For an arbitrary Markov branching process, we can use the formal decomposition 
$\A=\U+\R$
of its first-moment generator to build a parallel model with the same $\A$.
Since the theorem holds for the latter process, it also holds for the former;
this is some kind of ``invariance principle''. 
All that is lost is the \emph{probabilistic interpretation} given in the previous comment; 
such an interpretation may be regained with the help of the
size-biased tree construction of Subsec.~\ref{subsec:sizebias}, but
is then more involved.

\subsection{Reversible mutation rates, and symmetrization}
\label{subsec:reversibility}

We will now discuss the important special case that $\U$ is \emph{reversible},  
in that $\rho_iU_{ij}=\rho_j  U_{ji}$ for all $i,j\in S$. 
This is assumed in most models of nucleotide evolution, see, e.g.,
\cite[Ch.~13]{EwGr05}.
The interest in this case comes from the following facts.

\emph{\thesubsection.1 Explicit form of the rate function}.
For reversible $\U$, the maximization in \eqref{eq:I_U} can be carried
out explicitly, so that the rate function takes the closed form
\cite[p.~50, Ex.~IV.24]{deHo00}
\be{eq:Diriform}
 I_{\U}(\nu) = 
  - \Big \langle \text{\small$\sqrt{\frac{\nu}{\rho}},\, \U  
  \sqrt{\frac{\nu}{\rho}}$ } \Big \rangle_{\!\rho}\,;
\ee
here both the square root and the
fraction are to be read componentwise, and $\langle u,v \rangle_{\rho}$ 
denotes the Dirichlet form $\sum_i u_i v_i \rho_i$
for vectors $u,v$, and $\rho$. 
(It is an interesting fact that \emph{no} such simplification exists
for reversible Markov chains in \emph{discrete} time.)
Noting that $\rho_i>0$ for all $i \in S$ by irreducibility,
using the reversibility in the form
$\sqrt{\rho_i/\rho_j} \, U_{ij} = \sqrt{U_{ij} U_{ji}}$,
and recalling that $U_{ii} = - \sum_{j:j \neq i} U_{ij}$,
one readily finds that Eq.~\eqref{eq:Diriform} is equivalent to
\be{eq:Iroot}
  I_{\U}(\nu)= \frac{1}{2} \sum_{i,j \in S: \, i \neq j}
               \big(\sqrt{\nu_i U_{ij}} - \sqrt{\nu_j U_{ji}} \big)^2.
\ee

\emph{\thesubsection.2 Estimation of the reproduction rate from the
ancestral distribution}. The reversibility of $\U$ immediately implies that the
vector $\rho h:=(\rho_i h_i)_{i\in S}$ is a left eigenvector of $\A=\U+\R$ for
the principal eigenvalue $\lambda$, cf. \cite{BBBK05}.
Hence $\pi=\rho h$ up
to a normalization factor, and therefore $\alpha=\rho h^2$, or
$h=\sqrt{\alpha/\rho}$, again up to a normalization factor. (As before,
the square root and the fraction are to be read componentwise.)
This in turn means that $\alpha$, together with $\rho$, determines
the reproduction rate $R$ up to an additive constant. Indeed, suppose
that $R$ and $R'$ are two reproduction rates (for the same mutation
matrix $\U$) having the same ancestral distribution $\alpha=\alpha'$.
Then $h=h'$, whence $(R-R')h = (\lambda-\lambda')h$. As $h$ is strictly
positive, it follows that all components of $R-R'$ agree.

\emph{\thesubsection.3 Symmetrized mutation rates}.
For reversible $\U$, one can introduce the matrix
$\tA := (\tilde A_{ij})_{i,j \in S}$ by $\tilde A_{ij}= \sqrt{\rho_i} A_{ij} / \sqrt{\rho_j}$,
which is symmetric and has the same spectrum as $\A=\U+\R$.
The maximum principle of Thm.~\ref{thm:fundamental_eq} can therefore also be 
derived from the Rayleigh-Ritz (or Courant-Fisher)
variational principle for the leading eigenvalue of $\tA$;
see  \cite[Sec.~2]{BBBK05}. We emphasize,
however, that the large deviation approach to \eqref{eq:varprinc}
is not tied to reversible matrices and, as we have shown above, admits a natural interpretation in terms of the underlying family tree.
Nevertheless, we will take advantage of the symmetrization $\tA$ 
in Sect. \ref{sec:asymptotics} below. In particular, we will use
the (unique) decomposition $\tA = \F + \E$ into a symmetric Markov generator
$\F=(F_{ij})_{i,j \in S}$, defined through
\be{eq:F}
F_{ij} = \begin{cases} \sqrt{U_{ij} U_{ji}} = F_{ji} & 
                          \text{for } i \neq j, \\
 - \sum_{k \in S \setminus \{i\}} \sqrt{U_{ik}U_{ki}} & \text{for } i = j\,,
         \end{cases}
\ee
and a diagonal matrix $\E:= \text{diag}(E_i \mid i \in S)$
with elements\footnote{The corresponding equation
in \cite[Sec.~2]{BBBK05}, namely, the second-last equation on p. 88, 
is erroneous and should be corrected accordingly.
}
\be{eq:E}
E_i := \sum_{j \in S} \tilde A_{ij} =R_i + U_{ii} - F_{ii}=R_i+\sum_{j\in S}\sqrt{U_{ij}U_{ji}}\,.
\ee

\section{Unfolding the variational principle}
\label{sec:contract}

As we have seen, the maximum principle of Thm.~\ref{thm:fundamental_eq} 
provides some general insight into the competition,
and resulting tradeoff, between
mutation and reproduction. In general, however, it can not be solved explicitly.
This is because both the
maximization over the space $\sP(S)$, and the eigenvalue equations  determining
$\pi$, $h$ and thus~$\alpha$, are $|S|$-dimensional, and $S$ is
typically large. It is thus natural to ask whether one can obtain a  low-dimensional variational principle in a specific setting. In the rest of this paper we will therefore
confine ourselves to genetic \emph{models of sequence type} where each type  is specified by a sequence of letters from a finite alphabet.
The variational problem can then be split into two simpler ones,
a constrained variational principle with fixed mean letter composition, and a
maximization over all possible constraints. In some cases, each of  these two subproblems may be treated explicitly or, at least,  approximately.

\subsection{Lumping of sequence types, or: Choice of a type space}
\label{subsec:lumping}

As previewed in Subsection \ref{sec: SequenceSpace}, we will now
assume that the type of an individual is characterized by a  sequence of
letters from some finite alphabet
$\varSigma$, which leads to the type space $\varSigma^N$. If we assume
that the mutation and reproduction rates are invariant under permutations of sequence sites, as we will do in what follows, this sequence space can be lumped into the smaller space 
\begin{equation}\label{eq:simplex_S}
S=\Big\{ i \in \ZZ^{d} \mid i_{\ell} \geq 0 \: \text {for} \: \ell \in \varSigma, \;
\sum_{\ell \in \varSigma} i_{\ell} = N\Big \}\,,
\end{equation}
recall Fig.~\ref{fig:lumping}.
For example, this is possible for sequence space models with parallel mutation and reproduction, in which
\begin{enumerate}
\item[(L1)] all sites
mutate independently and according to the same (Markov) process
(a natural first assumption made in many models of sequence  evolution) and
\item[(L2)] the fitness function is invariant under permutation of sites
(a less natural, but still common assumption that applies,
for example, if fitness only depends on the sequence through
the number of mutated positions 
(i.e., the Hamming distance) relative to a
reference sequence, often termed the `wildtype');
\end{enumerate}
see, e.g., \cite{HRWB02,GaGr04} or 
\cite{GeHw02} for previous work on this  case.
(As an alternative to the choice \eqref{eq:simplex_S}, one can
use the constraint $\sum_{\l \in \varSigma} i_{\l} = N$ to remove an
element $a\in\varSigma$ by setting $ \varSigma^*= \varSigma\setminus\{a\}$
and work instead with
\be{eq:simplex_S*}
S^*=\Big\{ i \in \ZZ^{d} \mid i_{\ell} \geq 0 \: \text{for} \: \l \in  \varSigma^*,\;
\sum_{\l \in \varSigma^*} i_{\ell} \leq N \Big\}\,,
\ee
where now $d=\lvert \varSigma^* \rvert=\lvert \varSigma \rvert-1$.)

Specifically, if the reproduction and mutation rates on
$\varSigma^N$ are given
by  $\cR_{\sigma}$ and $\cU_{\sigma \tau}$ ($\sigma, \tau \in
\varSigma^N$), then, by permutation invariance,  there
is a vector $R=(R_i)_{i \in S}$ and a Markov generator
$\U=(U_{ij})_{i,j\in S}$ so that $\cR_{\s} = R_{H(\s)}$ and
$\sum_{\t: H(\t)=j} \cU_{\s \t} = U_{H(\s),j}$ for all $\s \in  \varSigma^N$;
here $H$ is as in \eqref{eq:H}.
$R$ and $\U$ then define a Markovian branching process with type  
space $S$.\footnote{
For a general description of lumping in Markov chains see
\cite[Ch.~6]{KeSn81}; and for an extension to the present (branching)
context with specific applications to genetics, see
\cite[Sec. 5 and 6]{BBBK05}. In the present case,
lumping is so immediate that it  hardly needs
to be formalized. But the procedure becomes nontrivial if, for example,
fitness functions are derived from Hopfield energy functions
(see \cite[Sec.~6]{BBBK05} and \cite{Gars05}).}

In fact, assumption (L1) even implies that the mutation rates
$U_{ij}$ of the lumped model are linear in $i\in S$ (or affine in $i \in S^*$).
This is seen as follows:
If $w_{\l m}$ is the mutation rate (at every site) from
letter $\l$ to letter $m$,
then the corresponding transition in the lumped model (based on
$\varSigma$) is  $i \to i - e_{\l} + e_{m}$ (where
$e_{j}$ is the the unit vector in $\R^d$ having a 1 at  coordinate $j$), 
and occur at rates
$i_{\ell} w_{\l m}$, due to independence of the sites.
If, instead, one removes one dimension by setting
$i_a =
N - \sum_{\ell\in\varSigma^*} i_{\l}$ and then works with
$\varSigma^*$, one obtains the additional
transitions $i \to i - e_{\l}$ at rate $w_{\l a} i_{\l}$, and
$i \to i +  e_m$ at the (affine)  rate
$w_{am} (N-\sum_{\ell\in\varSigma^*} i_{\ell})$.

Assumption  (L2) is less specific than (L1);
the fitness function in the
lumped model will,
in general, be nonlinear  due to interactions between sites.
It will, however, turn linear (or affine) if  fitness contributions
are additive across sites, as is usually assumed in, e.g., models of  codon bias
(where $\varSigma$ is the set of possible codons).
Additivity reflects independent fitness contributions of the sites
and means that, for $i \in S$, one has
$R_i = \sum_{\ell\in\varSigma} r_{\ell}\, i_{\ell}$
(based on $\varSigma$),
or $R_i =r_aN+\sum_{\ell\in\varSigma^*}^d (r_{\ell}-r_a) i_{\ell}$
(if   $\varSigma^*$ is used), where
$r_{\ell} \in \RR$ for $\ell \in \varSigma$.
We will examine such linear models in  Subsec.~\ref{subsec:linear}.

\subsection{Fixing the empirical mean}
\label{subsec:fixing_mean}

The only property of the special choices \eqref{eq:simplex_S} or
\eqref{eq:simplex_S*} of the type space~$S$
we need at the moment is that $S \subset \RR^d$. This provides
$S$ with the structure of an abelian group (elements of $S$ can be  added and subtracted), and allows us to classify the possible  empirical distributions
$\nu\in\sP(S)$ according to the value of their mean $\sp{\nu,\id} \in  \RR^d$.
In particular, for the
random measure $L^\omega(t)$ of \eqref{eq:L}, $\sp{L^\omega(t),\id}$ is
a random vector in $\RR^d$, namely the
\emph{empirical mean}, or \emph{empirical mean letter composition}
along the line $\omega$ up to time  $t$. 
If $S$ is obtained through lumping
a sequence space $\varSigma^N$ as in Fig. \ref{fig:lumping},
the $\l$'th coordinate of $L^ \omega(t)$
indicates the total fraction of time up to $t$ for which some site
in the sequence characterizing an individual on the line $\omega$ shows letter
$\l\in\varSigma$. Note that this involves a twofold averaging, namely  an average
over time and a (non-normalized) average over sequence sites.

As indicated in \eqref{eq:lambda_Lambda_nonstrict_preview} and
\eqref{eq:Lambda_preview}, we will now
foliate the variational formula \eqref{eq:varprinc} by prescribing the  mean
of the underlying type distribution. That is, we write
\be{eq:lambda_Lambda_nonstrict}
   \lambda = \max_{z \in \text{conv}\: S}  \Lambda(z),
\ee
where 
\be{eq:Lambda}
   \Lambda(z) :=
   \max_{\substack \nu \in \sP(S): \\ \langle \nu,\id \rangle = z} \;
   [\sp{\nu,R} - I_{\U}(\nu)]
\ee
is the \emph{constrained mean fitness} of $z\in \text{conv}\: S$.
As before, the maxima are attained by continuity, and the maximizer
in \eqref{eq:Lambda} is unique 
by the strict convexity of $I_{\U}$.
The function $\Lambda$ is strictly concave; this follows
again from the strict convexity of $ I_{\U}$, together with
the linearity of $\sp{\,\cdot\,,R}$ and $\sp{\,\cdot\,,\id}$.
In particular, $\Lambda$ is continuous on
\[
\text{rint\,conv}\: S=\big\{\sp{\nu,\id}\mid \nu\in \sP(S), \,\nu_i  >0 \text{ for all } i\in S\big\}\,,
\]
the  \emph{relative interior} of $\text{conv}\; S$
\cite[p.~82]{Rock70}. In general, the relative interior
$\text{rint}\: \sD$ of a set $\sD\subset\RR^d$ is defined as the  interior of $\sD$
relative to the smallest affine subspace containing $\sD$.%
\footnote{Recall that the simplex \eqref{eq:simplex_S} is contained  in a hyperplane,
so that the usual interior of its convex hull is empty.}
Moreover, since $\a$ is the unique maximizer in \eqref {eq:varprinc_alpha},
there exists a unique $ \hat z \in \text{conv\,}  S$
that maximizes $\Lambda$, namely
\begin{equation}\label{eq:anctype}
\hat z = \langle \alpha, \id \rangle,
\end{equation}
i.e., the unique maximizer $\hat z$ in \eqref {eq:lambda_Lambda_nonstrict}
is the ancestral type average.

If $\U$ is reversible, 
we may  restrict the maximization in \eqref{eq:lambda_Lambda_nonstrict}
to those $z$ that are \emph{strict} convex combinations of the 
elements of $S$. This  is obvious from the explicit form of $I_{\U}$
in \eqref{eq:Iroot}:  If at least one component of $\nu$ vanishes, one has
$(\partial/\partial \nu_i) I_{\U}(\nu)= + \infty$ for some $i$.
Therefore, the maximum will be located in $\text{rint\,conv}\: S$, so that
Eq. \eqref{eq:lambda_Lambda_nonstrict} can be replaced by
\be{eq:lambda_Lambda_strict}
   \lambda = \max_{z \in \text{rint\,conv}\,  S}  \Lambda(z).  
\ee

If the function $\Lambda(z)$ were known explicitly, the variational problem 
of Thm.~\ref{thm:fundamental_eq} would boil
down to a maximization over a subset of $\RR^d$; for small~$d$ one could
aim at explicit solutions. Such  
low-dimensional
variational principles for $\lambda$ were recently derived for several
examples, by methods from linear algebra  and asymptotic analysis
\cite{BBBK05,Gars05,GaGr04,HRWB02}. However, a 
plausible understanding for the resulting function to
be maximized has been lacking so far. 
The next Proposition reveals the probabilistic meaning of 
$\Lambda(z)$: It is nothing but the asymptotic growth rate of the lines having 
empirical type average (close to) $z$.
Together with \eqref{eq:anctype}, this shows that the growth rate of the total
population coincides with the growth rate $\Lambda(\hat z)$ of the subpopulation consisting of all individuals with empirical type average close to the ancestral one.
\begin{proposition}\label{prop:constrained_growth_rate}
For all $z\in \text{rint\,conv}\,  S$, the solution of the
constrained variational problem \eqref{eq:Lambda} satisfies
\[
\Lambda(z) = \lim_{\eps\to 0} \lim_{t \ti} 
     \frac{1}{t}   \log \EE^i \Big (\sum_{x \in X(t)}
     \II\big\{\| \langle L^x(t),
     \id \rangle - z \|_1 \leq \eps \big\} \Big ).
\]
\end{proposition}
\begin{proof}
Consider first the parallel model. By the reasoning leading to \eqref{eq:lambda_L},
the growth rate of the subpopulation consisting of all 
individuals with empirical mean close to $z$, up to some maximal 
deviation $\eps>0$, is equal to
\be{eq:empmean_z}
\begin{split}
  \lefteqn{\lim_{t \ti} \frac{1}{t}   \log \EE^i \Big (\sum_{x \in X(t)}
     \II\big\{\| \langle L^x(t),
     \id \rangle - z \|_1 \leq \eps \big\} \Big )}\\
& =   \lim_{t \ti} \frac{1}{t} \log   \sum_{n \geq 0} 2^n\; 
   \EE^i \Big ( \II\big\{\beta(t)=n,\;T>t,\; \|\sp{L(t),\id}-z\|_1 \leq \eps \big\}\Big ) \\ 
& =  \lim_{t \ti} \frac{1}{t} \log \EE^i \Big (
       \e^{t \langle L(t),R \rangle}\,
       \II \big\{\| \langle L(t),\id \rangle - z \|_1 \leq \eps  \big\} \Big )\\[.5ex]
  &   = \lim_{t \ti} \frac{1}{t} \log \EE^i \Big (
        \exp\big[\,t\,\big(\langle L(t),R \rangle
 -\infty \cdot \II\{\| \langle L(t),\id \rangle - z \|_1 > \eps \}\big)\big] \Big ) \\[.5ex]
   & = \max_{\nu \in \sP(S):\, \| \langle \nu, \id \rangle - z \|_1 \leq \eps }
      \big [ \langle \nu, R \rangle - I_{\U}(\nu) \big ]
      = \max_{y\in\text{conv}\,  S:\, \| y- z \|_1 \leq \eps }\Lambda(y). 
\end{split}
\ee
Here we have used the conventions $\infty \cdot 1 =\infty$ and
$\infty \cdot 0 =0$ in the third step, and 
Varadhan's lemma in the fourth, in analogy with \eqref{eq:varprinc};
the maximum over $\nu$ is attained since the condition
$ \| \langle \nu, \id \rangle - z \|_1 \leq \eps$ defines a
compact subset of $\sP(S)$.
As $\Lambda$ is continuous on $\text{rint\,conv}\,  S$,
the last expression converges to  $\Lambda(z)$ as $\eps \to 0$,
as asserted.

For a general splitting rule, the argument is the same except that
the particular tree construction of Subsec. \ref{subsec:parallel_tree}
has to be replaced by the size-biased tree described in
Subsec.~\ref{subsec:sizebias}. In fact, one simply has to omit the second line
of \eqref{eq:empmean_z} above and instead invoke Eq. (4.4) of \cite{GeBa03}
which shows that the first line of \eqref{eq:empmean_z} coincides with the third;
the random measure $L(t)$ in the third line is then 
again the empirical measure of a Markov chain with generator~$\U$,
namely the mutation process along the spine of the size-biased tree.
\end{proof}

Like the unconstrained variational problem \eqref{eq:varprinc}
leading to $\lambda$, the constrained 
problem \eqref{eq:Lambda} defining $\Lambda$
provides insight into the mutation-re\-pro\-duction process,
but does not, in general, lead to an explicit solution if $S$ is large.
>From the point of view of explicit calculations, it rather
expresses one difficult problem (the leading eigenvalue of
a large matrix) in terms of another difficult problem (the
maximization over a large space). 
But if $\U$ is reversible, there are two cases in which
\eqref{eq:Lambda} may be solved explicitly or, at least, asymptotically. 
These  are the cases when fitness and mutation are linear (already hinted at 
in Sec.~\ref{subsec:lumping}), or when they allow a continuous approximation 
in the limit as the number $N$ of sequence sites  grows large. 
These will be discussed in the next Subsection and in 
Section~\ref{sec:asymptotics}.

\subsection{Exact results for linear reversible models}
\label{subsec:linear}

In this Subsection we have a closer look at the sequence space models
of Sec.~\ref{subsec:lumping} that describe the
independent evolution of $N$ sites with a finite alphabet $\vS$
and lead, after lumping, to models with
state space $S$ as in \eqref{eq:simplex_S},
with linear fitness \emph{and} mutation, and
mutational transitions $i \to i+k$  restricted to those with
$k \in \fS :=\{e_m - e_{\ell} \mid m,\ell\in\vS, \,m\ne\ell\}$.
In line with standard  assumptions on sequence evolution (see, e.g.,
\cite[Ch.~14]{EwGr05}), we posit that the mutation process
acting at the sites is reversible, that is, the mutation rates $(w_ {\l m})_{\l,m\in\vS}$
define an irreducible and reversible Markov generator with a
reversible distribution $\gamma$ on $\vS$. After lumping, the associated
mutation process on $S$ then has rates
$U_{ij} = U_{j-i} (i)$ , $i,j\in S$, given by
\[
  U_k(z) = \begin{cases} w_{\ell m} z_{\ell} & \text{if } k=e_m - e_ {\ell}\in \fS, \\
                         - \sum_{\ell \neq m} w_{\ell m} z_{\ell} & \text{for } k=0, \\
                         0 & \text{otherwise}
           \end{cases}
\]
for $z\in\RR^d$, $d=|\vS|$.
The reversibility of $(w_{\l m})_{\l, m\in\vS}$ readily implies that  the mutation
generator $\U=(U_{ij})_{i,j\in S}$ is also reversible; its reversible  distribution is
$\rho:= \Mult_{N,\gamma}$\,, the multinomial distribution for $N$  samples from
the distribution $\gamma$ on $\vS$.
As motivated in Sec.~\ref{subsec:lumping},
we will also assume here that the reproduction rates are linear, in that
$R_i = R(i)$ for all $i \in S$, for a linear function $R$ of the form
$R(z) = r\cdot z$, $r,z\in\RR^d$.
Here and below we write `$\cdot$' for the scalar product of vectors in
$\RR^{d}$, in contrast to $\sp{.,.}$, which we have reserved for  scalar
products of vectors in $\RR^{S}$.
In this setting, the constrained variational problem \eqref{eq:Lambda}
admits an explicit solution as follows. Due to \eqref{eq:lambda_Lambda_strict},
we may -- and will -- restrict ourselves to considering
means in
\[
\text{rint\,conv}\: S
=\Big\{ z \in \RR^d \mid z_{\l}>0 \: \text{ for } \: \l \in  \varSigma, \;
\sum_{\l \in \varSigma} z_{\ell} = N \Big\}\,.
\]

\begin{theorem}\label{thm:Lambda_linear}
In the situation described above, for every $z \in \text{\rm rint\,conv}  \: S$
the restrained maximum of \eqref{eq:Lambda} is given by
\be{eq:Lambda_linear}
  \Lambda(z)   = R(z) - \frac12\sum_{k \in \fS}
  \Big ( \sqrt{U_k(z)} - \sqrt{U_{-k}(z)} \Big ) ^2
    = \langle \nu^{(z)}, R \rangle - I_{U}(\nu^{(z)}),
\ee
where $\nu^{(z)} =\Mult_{N,z/N}$ is the multinomial distribution with  mean $z$.
\end{theorem}

\begin{proof}
Let $z \in \text{rint\,conv} \: S$ be given, and consider
any $\nu\in\sP(S)$ with $\sp{\nu, \id} = z$.
It is then clear that $\sp{\nu, R}=R(z)$ by linearity.
Let us rewrite Eqn.~\eqref{eq:Iroot} in the form
$I_{\U}(\nu) = \frac12\sum_{k \in \fS}   
\| x_k -  y_k \|_2^2 $, where $x_k = (x_{k,i})_{i \in S}$ and
$y_k = (y_{k,i})_{i \in S}$ are the vectors with components
\[
   x_{k,i} = \sqrt{U_k(i) \nu_i}, \quad
   y_{k,i} = \sqrt{U_{-k}(i+k) \nu_{i+k}}\,,
\]
$i \in S$, $k \in \fS$.
In the boundary case when $i\in S$ but $i+k\notin S$, we have
$U_k(i)=0$ by definition, and likewise $U_{-k}(i+k) = 0$ when
$i\notin S$ but $i+k\in S$. Hence $ x_{k,i} =  y_{k,i}=0$ unless
$i,i+k\in S$. By linearity of the $U_k$,
it follows that
$\| x_k \|_2^2 = U_k(z)$ and
$\| y_k \|_2^2 = U_{-k}(z)$ for all $k \in \fS$.
As the distance between any two vectors is minimized when the vectors  are parallel,
we conclude further that
\[
\| x_k -  y_k \|_2^2   \geq \big ( \| x_k \|_2 - \| y_k \|_2 \big )^2
\]
with equality if and only if there is a positive constant
$C_k$ so that
\[
  U_k(i) \nu_i = C_k U_{-k}(i+k) \nu_{i+k}
\]
whenever $i ,i+k\in S$.
This, however, is the case when $\nu=\nu^{(z)}$
because, for each $i\in S$,
$\nu^{(z)}_i =e^{\b\cdot i}\rho_i$ for $\b=\log({z}/{N\gamma})$
(where the fraction and the logarithm are taken componentwise), and  $ \rho$ is reversible; in fact we have $C_k = \e^{-\b \cdot k}$.
Combining the preceding observations we get the result.
\end{proof}

  If we turn from the linear model to the affine one,
  by removing one coordinate as indicated at the close of
  Sec.~\ref{subsec:lumping}, Theorem \ref{thm:Lambda_linear} clearly remains 
  true, with the middle expression in \eqref{eq:Lambda_linear} expressed in
  terms of the reduced coordinates. 
  Eq.~\eqref{eq:Lambda_linear} has been derived previously for certain
  specific choices for the mutation rates \cite{GaGr04,HRWB02};
  remarkably, the above
  result provides both an extension (to arbitrary reversible models),
  and a simplification of the proof.

\subsection{Partial convex conjugation}
\label{subsec:general_case}

On our way to the second case of an explicit version 
of  $\Lambda(z)$, we need a general
intermediate step: a relation between $\Lambda(z)$ and the mean growth rate
$\lambda$ for a suitably modified reproduction rate $R$.
This relationship is based on partial convex
conjugation, a standard procedure of convex analysis which will
be spelt out here for our purposes. In Sec. \ref{sec:asymptotics}, 
this will allow us  to determine the asymptotic behaviour 
of $\Lambda(z)$ when the number $N$ of sequence sites gets large.

Let us rewrite equation \eqref{eq:varprinc} in the form 
\[
\lambda(R) = \max_{\nu \in \sP(S)} \big[ \langle \nu,R \rangle - I_{\U}(\nu)\big]
\] 
indicating the dependence on $R$; $\U$ will be considered as fixed. 
The following proposition asserts that the function $z\to-\Lambda(z)$
of  constrained extrema is a partial convex conjugate of the function 
$R\to\lambda(R)$.
\begin{proposition}\label{prop:Lambda_general}
Let $S \subset \RR^d$, $z \in \text{\rm rint\,conv\,} S$,
and $\U$ be an irreducible Markov generator on $S$
(not necessarily reversible). Then the constrained variational
problem \eqref{eq:Lambda} has the solution
\[
\begin{split}
\Lambda(z) & =  \inf_{\beta \in \RR^d}  \big [\lambda(R+\beta \cdot \id)
        - \beta \cdot z   \big ] \\
        &= \lambda(R+\beta^z \cdot \id) - \beta^z\cdot  z
         =  \langle \alpha^z ,R  \rangle 
          - I_{\U}( \alpha ^z )\,.
\end{split}
\]
Here, $\beta^z \in \RR^d$ is the negative slope vector of any tangent plane
to $\Lambda$ at $z$, 
and $\alpha^z$ is the unique ancestral distribution corresponding to the
reproduction rate $R+\beta^z \cdot \id$ for any such 
$\beta^z$. In particular, the function $\beta\to \lambda(R+\beta\cdot\id)$ 
is differentiable on $\RR^d$, and $\nabla_{\!\beta\,}
\lambda(R+\beta\cdot\id)\mid_{\,\beta=\beta^z}\,=\sp{\alpha^z,\id}=z$.
\end{proposition}

\begin{proof}For any $z\in \text{rint\,conv\,} S$ and $\beta \in \RR^d$ we have, 
writing $\nu^z$ for the maximizer in \eqref{eq:Lambda} and using
Thm.~\ref{thm:fundamental_eq}, 
\[
\Lambda(z) = \sp{\nu^z,R+\beta\cdot\id}-\beta\cdot z-I_{\U}(\nu^z)\le
\lambda(R+\beta\cdot\id)-\beta\cdot z\,.
\]
Taking the infimum over $\beta$ we arrive at the inequality
\begin{equation}\label{eq:Lambda_ineq}
\Lambda(z) \le  \inf_{\beta \in \RR^d} [\lambda(R+\beta\cdot\id)-\beta\cdot z]\,.
\end{equation}
To show equality we recall that $\Lambda$ is strictly concave 
and finite on a (relative) neigbourhood of $z$ and therefore admits a tangent plane at $z$.
That is, there exists some $\beta\in \RR^d$ such that
\[
\Lambda(y)\le \Lambda(z)-\beta\cdot (y-z) \quad\text{ for all }y\in\text{conv } S\,,
\]
with strict inequality for $y\ne z$.
Denoting by $\alpha^\beta$ the ancestral distribution for the reproduction rate
$R+\beta\cdot\id$ and letting $y=\sp{\alpha^\beta,\id}$ we find
\begin{equation}\label{eq:Lambda_ineq2}
\begin{split}
\lambda(R+\beta\cdot\id)-\beta\cdot z&=
\sp{\alpha^\beta,R}+\beta\cdot(y-z)-I_{\U}(\alpha^\beta)\\
&\le\,\Lambda(y)+\beta\cdot(y-z)\,\le\, \Lambda(z)\,.
\end{split}
\end{equation}
Together with \eqref{eq:Lambda_ineq} it follows that equality holds
everywhere in \eqref{eq:Lambda_ineq2}. Hence $y=z$, \eqref{eq:Lambda_ineq} holds
with equality, and the infimum is attained for any $\beta$ determining a tangent
to $\Lambda$ at $z$. In general, there may be several such
tangents,  e.g., if $S$ is contained in a hyperplane of $\RR^d$.
However, the associated ancestral distribution is uniquely determined.
For, suppose there exist $\beta_1\ne\beta_2$ both determining a tangent to $\Lambda$ at $z$,
and let $\alpha^1$ and $\alpha^2$  be the ancestral distributions for the reproduction
rates $R+\beta_1\cdot\id$ and $R+\beta_2\cdot\id$, respectively.
The preceding argument then holds for every $\beta$ in the segment
$[\beta_1,\beta_2]$, whence  \eqref{eq:Lambda_ineq2} holds with equality everywhere
for all these $\beta$. We  can thus conclude
that the function $\beta\to\lambda(R+\beta\cdot\id)$ is affine on 
$[\beta_1,\beta_2]$.
In particular, using Thm.~\ref{thm:fundamental_eq} and the shorthand 
$f_i(\nu)=\sp{\nu,R+\beta_i\cdot\id}-I_{\U}(\nu)$  we find
\[
\max_\nu\big[\frac12 f_1(\nu)+\frac12 f_2(\nu)\big] =
\frac12\max_\nu f_1(\nu)+\frac12 \max_\nu f_2(\nu).
\]
Since $f_1$ and $f_2$ are strictly concave, this is only possible 
if they have the  same maximizer. 
That is, $\alpha^1=\alpha^2$.
Finally, using the equality in \eqref{eq:Lambda_ineq} and the 
convex duality lemma \cite[Lemma 4.5.8]{DeZe98}
we find that the function $\beta\to \lambda(R+\beta\cdot\id)$ is the
convex conjugate of the strictly convex function $-\Lambda$, and thus
differentiable; see \cite[Thm.~26.3, p.253]{Rock70}.
Its gradient at $\beta^z$ necessarily coincides with $z$.
\end{proof}

In the case of a reversible mutation matrix $\U$, the preceding
proposition can be complemented as follows.
We write $T=\text{\rm span}(S-S)\subset\RR^d$ for 
the linear space generated by the set of differences of elements of $S$.
\begin{corollary}\label{cor:Lambda_rev} 
For reversible $\U$ the following additional statements hold.\\
(a)  The function $\Lambda$ defined in \eqref{eq:Lambda} is differentiable%
\footnote{If $T$ is a proper subspace of $\RR^d$, differentiability means
that the directional derivatives in the directions of $T$ exist, and the gradient is the unique element of $T$ determined by these directional derivatives; its component orthogonal to $T$ is thus set equal to zero.}%
on $\text{\rm rint\,conv\,} S$, and its conjugate function $\beta\to
\lambda(R+\beta\cdot\id)$ is strictly convex on $T$. 
Moreover, 
for $z\in\text{\rm rint\,conv\,} S$ and $\beta\in T$ we have $\beta=-\nabla\Lambda(z)$
if and only if $z=\nabla_{\!\beta\,}\lambda(R+\beta\cdot\id)$.\\
(b) The function $\Lambda$ on $\text{\rm rint\,conv\,} S$ remains unchanged 
under symmetrization, i.e., by
replacing $\U$ with the matrix $\F$ of \eqref{eq:F}, and $R$ with the function
$E$ defined in \eqref{eq:E}.
\end{corollary}
\begin{proof}
(a) Let $z\in\text{rint\,conv\,} S$ and $\beta,\beta'$ be
two negative slope vectors of $\Lambda$ at $z$.
In view of the uniqueness of $\alpha^z$ and 
the remarks in \ref{subsec:reversibility}.2,
the scalar product $(\beta-\beta')\cdot i$ is then independent of $i\in S$. 
This means that $\beta-\beta'$ is orthogonal to $T$, so that there is a unique
negative slope vector $\beta^z\in T$. By concavity, the uniqueness
of the tangent plane is equivalent to differentiability; cf.\
\cite[Thm.~25.1, p.~242]{Rock70}.
By the proof of Prop.\ \ref{prop:Lambda_general},
this is also equivalent to strict convexity of $\lambda(R+\beta \cdot \id)$ on $T$. 
The final statement comes from the observation that both assertions are equivalent
to the identity $\lambda(R+\beta\cdot\id)-\Lambda(z)=\beta\cdot z$.

(b) For each $\beta\in \RR^d$, the matrix $\F+\text{diag}(E_i+\beta\cdot i\mid i\in S)$
is similar to $\U+\text{diag}(R_i+\beta\cdot i\mid i\in S)$, so that their principal 
eigenvalues agree. The result thus follows 
from Prop.\ \ref{prop:Lambda_general} by minimization over $\beta$.
\end{proof}

\section{Smooth approximations}  
\label{sec:asymptotics}

While still adhering to a lumped sequence model,
we will now turn to a situation complementary to that of Thm.~\ref{thm:Lambda_linear}:
we consider \emph{nonlinear}
reproduction and mutation rates that allow for a 
continuous approximation
if the number of sequence sites becomes large; this
approximation is only required {locally}, which provides much more
freedom, and, in particular, removes constraints imposed by the
boundary (recall the boundary conditions $U_k(i)=0$ for $i \in S$, $i+k \notin S$
in Thm.~\ref{thm:Lambda_linear}). 
For a large family of models with \emph{reversible} $\U$, 
an asymptotic low-dimensional
maximum principle for $\lambda$ is available then \cite{BBBK05},
but no connection to the constrained mean fitness \eqref{eq:Lambda} has been
made there, and the ensuing probabilistic interpretation was still lacking.
On the basis of Prop.~\ref{prop:Lambda_general}, this can now be provided.

In view of Corollary \ref{cor:Lambda_rev}(b), the case of a reversible 
mutation matrix $\U$ can be reduced to the case of a symmetric
mutation matrix $\F$. That is, instead of the first moment generator $\A=\U+\R$
we can and will consider the symmetrized version $\widetilde \A= \E + \F$
defined in \eqref{eq:F} and \eqref{eq:E}. In Subsection~\ref{subsec:approx_lambda} 
we will present a slight refinement of an asymptotic maximum principle derived in
\cite[Thm.\ 1]{BBBK05}. In  Subsection \ref{subsec:Lambda} we will derive an
approximation of $\Lambda(z)$ in two  particularly interesting situations.
An application to the quasispecies model follows in Sec. \ref{sec:quasispec}. 

\subsection{Approximation of the asymptotic growth rate $\lambda$}
\label{subsec:approx_lambda}

Consider the following setup. For each $N$ let
\begin{itemize}
\item 
$S=S(N)\subset\ZZ^d$ be a state space as in \eqref{eq:simplex_S} or 
\eqref{eq:simplex_S*}.
\end{itemize}
The rescaled set $\frac{1}{N} S$ is then contained in a simplex $\sD\subset\RR^d$, 
viz. either $\sD= \text{conv}\: \{e_1, \ldots, e_d \}$ or 
$\sD= \text{conv}\: \{0, e_1, \ldots, e_d \}$,
with $e_1, \ldots, e_d$ the unit vectors of $\ZZ^d$.
(In the first case, $\sD$ is contained in a hyperplane, whence in the following we will
always consider the \emph{relative} interior of $\sD$ rather than simply its interior.)
In the limit as $N\ti$, $\frac{1}{N} S$ becomes dense in $\sD$.
For each $N$ let also
\begin{itemize}
   \item $ \F$ be a symmetric Markov generator  on $S$, and 
        $\E:= \text{diag}(E_i \mid i \in S)$ a diagonal matrix.
\end{itemize}
We assume that $\F$ and $\E$ admit a continuous approximation 
as follows: There exist real functions 
$e$ and $f_k$ on $\sD$, and
an ``approximation domain'' $\sA\subset\text{rint}\,\sD$
such that the following conditions hold.
\begin{description}
   \item[\rm(A1)] $e$ is $C^2$
   on $\sA$ and, as  $N \ti$,
\[
   E_i=e  \biN + \cON
   \quad \text{and} \quad
  F_{ij} =f_{j-i} \biN  + \cON \,,
\]
where the $\cO(1/N)$ terms
are uniform for all $i,j\in S$ with $i/N, j/N\in\sA$.
   \item[\rm(A2)]
Uniformly for all $i$ with $i/N\in\sA$, 
\[
    \sum_{k \in S-i} f_k \biN \,
    \lvert k^{}_{\ell} \rvert k_m^2 \leq C
\]
for some constant $C$ and all $1 \leq \ell,m \leq d$, 
where $S-i := \{j-i : j \in S\}$.
   \item[\rm(A3)] For suitable constants $C', C''<\infty$ we have
\[
-C' \leq E_i \leq \sup_{z\in\sD} e(z) + \cON \quad\text{and}\quad F_{ij} \leq C'' 
\]
for all $i,j \in S$, $i \neq j$, with a uniform error term $\cO(1/N)$.
\end{description}

\begin{theorem}\label{thm:lambda_asym}
Suppose the conditions  (A1)--(A3) hold for a relatively open neighbourhood
$\sA$ of a global maximizer $z^*\in\text{rint}\,\sD$ of $e$. Then
the principal eigenvalue $\lambda$ of the matrix $\tA = \E + \F$
admits the approximation
\[
    \lambda  = e(z^*) + \cON.
\]
The error term here only depends on the constants in (A1)--(A3) and the
Hessian of $e$ at $z^*$ (via an upper bound on the modulus of its
most negative eigenvalue).
\end{theorem}
We postpone the proof until Subsection \ref{sec:proofs}, discussing first
the significance of the assumptions and the result.

\emph{\thesubsection.1 Formal comments}.
The above approximation for the principal eigenvalue of $\tA$
clearly also holds for the similar matrix $\A=\R+\U$.
Note also that only the function $e$ 
remains relevant in the limit; the $f_k$ play no role. This means
that $\tA=\E+\F$ provides  the `right' decomposition into the `relevant'
$\E$-term, and an $\F$-term whose contribution to  the leading 
eigenvalue vanishes in the limit. 

It is also interesting to observe that the approximation assumption
(A1) is only required in a  neigbourhood $\sA$ of a
single maximizer $z^* \in \text{rint\,} \sD$ of $e$;
further maxima may appear, even on the boundary, but these do not matter. 
This locality of the approximation domain is the main difference to
Thm.~1 of \cite{BBBK05} which requires a globally uniform
approximation. 
(As the example of linear mutation in Thm.~\ref{thm:Lambda_linear} shows,
it often happens that the derivatives diverge at the relative boundary of $\text{conv\,}S$,
so that a global approximation is not feasible. This is also the case for the
quasispecies model considered in Sec. \ref{sec:quasispec}.)
As a global requirement we need only the bounds in (A3).

\smallskip
\emph{\thesubsection.2 Significance of the assumptions for the model}.
Our setup implies that replacing  $i \in S$  by $i/N \in S/N$
will yield a continuous type variable $z \in \sD$ in the limit.
Accordingly, the matrix elements are required to become smooth 
functions of $z$ as $N \ti$ -- at least locally, in line with (A1). 

Condition (A2) says that the mutation rates must decay fast enough
with distance to the target type -- again, at least locally.
This assumption may appear to be rather special at first sight, but 
actually it is very natural:
As we have seen in Sec.\ \ref{subsec:lumping},
independent mutation at the sites of a sequence
leads to nearest-neighbour mutation  on $S$, hence
(A2) is  trivially fulfilled. For the corresponding quasispecies model
(to be described below), still with independent mutation at the sites,
the decay of $f_k$ with $k$ is  exponential, rather than
only cubic as required in (A2);
this will be shown in Sec.~\ref{sec:quasispec}.

In many concrete examples, the reproduction and mutation 
rates have their own continuous approximations each, i.e., 
\[
R_i = r \biN + \cON \quad\text{ and }\quad U_{ij} = u_{j-i} \biN + \cON
\] 
with $C^2(\sD,\RR)$ functions
$r$ and $u_k$. Moreover,  the range of
all mutational steps is finite (on $S$, and independently of $N$); that
is, there is a 
finite symmetric (i.e., $\fS = - \fS$) set $\fS \subset \ZZ^d$ with
the property that, for all $N$, $ U_{ij}=0$ 
whenever $j-i \notin \fS$. Then (A1)  is
automatically satisfied for any $\sA$ on which  $\sqrt{u_{k}(z)}$ 
is $C^2$  for all $k \in \fS$;  inspecting 
the matrix elements of $\E$ in \eqref{eq:E} and noting that 
$\sum_{k \in \fS} u_k(z)=0$ one finds that 
\be{eq:e_parallel}
e(z) =  r(z) - \frac12\sum_{k \in \fS} 
  \big ( \sqrt{u_k(z)} - \sqrt{u_{-k}(z)}\, \big ) ^2. 
\ee
It is interesting to observe that the expression above
is formally identical with $\Lambda(z)$ of 
Thm.~\ref{thm:Lambda_linear},
although we are considering quite a different situation here. 
Special cases of \eqref{eq:e_parallel} have appeared in 
\cite{BBBK05,HRWB02} in the context of parallel sequence space models,
and the resulting maximum principle turned out as a key to determine
the mutation load, genetic variance,
and the existence of error thresholds. 

\emph{\thesubsection.3 Locality of the ancestral distribution}.
Under the additional (but generic) assumptions
that the function $e$ admits a \emph{unique}
maximizer $z^* \in \text{rint}\,\sD$ and the Hessian of $e$ at $z^*$,
restricted to $T=\text{span}(\sD-\sD)$,
is (strictly) negative definite,  one can also characterize the
ancestral distribution, which is connected to $\lambda$ 
through the general variational principle of Thm.~\ref{thm:fundamental_eq}.
Namely, by Thm.~2 of \cite{BBBK05}, this distribution
is concentrated in a neighbourhood of $z^*$ whose width decreases
with $1/\sqrt{N}$. More precisely: 
For every  $0 < \eps \leq 1$, there is a
constant $c > 0$, independent of $N$, so that, for $N$ large enough,
\be{eq:sqrt_ancestors}
  \sum_{\substack {i \in S: \\   \lvert i/N - z^* \rvert \geq c / \sqrt{N}}}
  \!\!\!\!\! \a_i \ \leq \eps\,.
\ee
By Cor.~3 of \cite{BBBK05}, it follows that
$\langle \alpha, \id \rangle = z^* + \cO ( N^{-1/3}  )$,
i.e., the ancestral type average coincides with the unique maximizer of $e$
up to a small error term. 
The constant in the error term here depends on those in the assumptions
and some bounds separating the spectrum of the Hessian (restricted to $T$)
 of $e$ at $z^*$ from $-\infty$ and $0$.
The proofs given in \cite{BBBK05} are solely based on a \emph{local}
approximation and thus remain valid under our weaker
assumptions.

\subsection{Approximations of the constrained mean fitness $\Lambda$}
\label{subsec:Lambda}

Our next goal is an approximation for the partial maximum $\Lambda(z)$ of 
(\ref{eq:Lambda}). In fact, 
the similarity of the expression \eqref{eq:e_parallel} for 
 $e(z)$ and the expression for $\Lambda(z)$ in
Thm.~\ref{thm:Lambda_linear} leads one to ask whether the asymptotic
identity of the global maxima of $\Lambda$ and $e$, as asserted by Theorem 
\ref{thm:lambda_asym}, can be extended to an asymptotic relation 
between these functions as a whole.  On the basis of Prop. \ref{prop:Lambda_general}
such an approximation can indeed be given. We consider first the most salient points of $e$,
i.e., the points where $e$ coincides with its concave envelope.  
Let us say $z\in \text{rint}\,\sD$ is an \emph{exposed smoothness point} of $e$ if
\begin{itemize}
\item[--] $e(y)<t_z(y):=e(z) +\nabla e(z)\cdot (y-z)$  for all $y\ne z$,
i.e., $z$ is the unique point where $e$ hits its tangent plane $t_z$ at $z$.
\item[--]$e$ is $C^2$ on a neighbourhood of $z$, and 
the Hessian of $e$ at $z$, as a bilinear form on $T=\text{span}(\sD-\sD)$, 
is negative definite.  
\end{itemize}
If $e$ is strictly concave on $\sD$, the first condition is trivially
satisfied. If $e$ is also $C^2$, the second condition just covers the generic case
of strict concavity. In other words, for a generic strictly concave $C^2$-function $e$,
every $z\in\text{rint\,}\sD$ is an exposed smoothness point. 

To state the hypotheses of the next theorem we recall that the assumptions
(A1) and (A2) only involve an approximation on a local set $\sA$,
while (A3) imposes some global bounds, including an upper bound on $E_i$ in terms of $\sup e$.
We now replace the  constant $\sup e$  by suitable tangent planes of $e$, thereby
turning (A3) into the hypothesis
\begin{description}
 \item[\rm(A3$'$)] For all $z\in\sA$ and suitable constants $C', C''<\infty$ we have
\[
-C' \leq E_i \leq t_z\Big(\iN\Big) +  \cON \quad\text{and}\quad F_{ij} \leq C'' 
\]
for all $i,j \in S$, $i \neq j$, with a uniform error term $\cO(1/N)$.
\end{description}
\begin{theorem}\label{thm:Lambda_e}
Consider a relatively open convex subset $\sA$ of $\text{rint}\,\sD$ 
consisting of exposed smoothness points of $e$
and satisfying the hypotheses (A1), (A2), and (A3\,$'$). Then 
one has the approximation 
\[\Lambda(z) = e(z)+\cO(N^{-1/3})\]
locally uniformly for all $z\in\sA$. The constants in the error term only depend
on the error terms in the assumptions, some locally uniform upper bounds on $|\nabla e|$, 
and the Hessian of $e$ (via some locally uniform  bounds separating its spectrum
from $-\infty$ and $0$).
\end{theorem}
The proofs of this and the subsequent theorem follow in the next subsection.

Theorem \ref{thm:Lambda_e} raises the question of what happens if $e$ 
touches a tangent plane at
two or more distinct points of $\text{int\,}\sD$. Let $z$ be a strict convex 
combination of these points, and $\beta$ the negative slope of  this plane.
The ancestral distribution $\bar\alpha$ for the reproduction rate 
$\bar E = E +\beta\cdot\id/N$ is then expected to split into 
distinct peaks located at the competing maximum points  of the associated $\bar e$; its mean type
$\sp{\bar\alpha,\id}$ will remain close to $z$, but the reproduction rate 
$\Lambda(z)$ will be the corresponding convex combination 
of the values of $e$ at the maximum points of $\bar e$. So it may be conjectured
that, in general, $\Lambda(z)$ is approximated by the 
concave envelope of $e$ at $z$.  The next theorem shows that this is indeed the case.
We note that this kind of  behaviour is related to
the phenomenon of error thresholds and phase transitions described
in detail in \cite{HRWB02}.

For a given function $e$ on $\sD$ we let
\[
\hat e(z) := \inf \Big\{a-\beta\cdot z \,\Big|\,  a\in \RR, \beta\in\RR^d, \ 
a -\beta\cdot y \ge e(y) \ \forall\,y\in\sD \Big\}\,,\quad z\in\sD,
\]
be the \emph{concave envelope} of $e$. 
(For an example see Fig.~\ref{fig:quasispec}.)
We consider the situation when $e$ deviates from strict concavity,
so that $\hat e$ is affine on a nontrivial set $\sB$. Let us say that $\sB$ 
is a \emph{basin}  of $e$ if  $\sB$ has nonempty relative interior and 
\begin{equation}\label{eq:basin}
\sB=\{z\in \sD\mid \hat e(z) =a-\beta\cdot z\}
\end{equation}
for suitable $a\in\RR, \beta\in\RR^d$. Note that a basin $\sB$ is 
necessarily convex and compact.
We write $\text{ex\,}\sB$ for the set of its extremal points.
Let us say that a basin $\sB$ of $e$ is \emph{determined by smooth hills} 
of $e$ if there exists 
a relatively open neigbourhood $\sH$ of $\text{ex\,}\sB$ in $\sD$ such that 
\begin{itemize}
\item[--] $\sH\setminus\sB$ consists of exposed points of $e$, and
\item[--] $e$ is $C^2$ on $\sH$, and its Hessian (restricted to $T$) is negative definite with 
a spectrum which is bounded from below and bounded away from zero uniformly on $\sH$.
\end{itemize}
This is the situation one typically encounters when a smooth $e$ deviates 
from strict concavity.
The theorem below provides an approximation of the restrained maximum 
$\Lambda(z)$
defined in \eqref{eq:Lambda}. 
\begin{theorem}\label{thm:basin}
Consider a basin $\,\sB\subset\text{rint}\,\sD$ of $e$ that is determined by smooth hills 
$\sH$ of $e$, 
and suppose the assumptions (A1), (A2), and (A3\,$'$) are satisfied
with $\sA=(\sH\setminus\sB)\cup\text{\rm ex\,}\sB$.
Then we have the approximation 
\[\Lambda(z) = \hat e(z)+\cO(N^{-1/3})\]
uniformly for all $z\in\sB $.
\end{theorem}

\subsection{Proofs}
\label{sec:proofs}

We now turn to the proofs of the three Theorems of this section.
\begin{proof}[of Thm.~\ref{thm:lambda_asym}]
The proof of Thm.~1 of \cite{BBBK05} goes through with the  changes
summarized below; we will refer to equations in the previous paper 
by double brackets $((.))$.
Throughout, notation changes from  $x$  to $z$,  $E(x)$ to  $e(z)$,
and $\alpha$ to $a$; $\text{int}\: D$ is replaced by 
$\text{rint}\: D$ throughout. 
The upper bound on $\lambda$ remains
unchanged in view of ((12)) and (A3). For the lower bound,
let $z^*$ 
be given as required, 
and place the test function 
$v=(v_i)_{i \in S}$ of ((32)) 
at this $z^*$. The argument after ((40)) changes as follows.
Due to (A1), $\exists \, 0 < \delta \leq \eps$ and 
$0 \leq \gamma < \infty$ so that, for $\lvert z - z^* \rvert < \delta$,
$e(z) \geq e(z^*) - \gamma \lvert z - z^* \rvert^2$. Then one has
\begin{eqnarray*}
    \sum_{i\in S}v_i^2 E_i 
  & = & 
    \sum_{\substack{i \in S: \\ \lvert i/N - z^* \rvert < \delta}}
    E_i \, v_i^2  
  + \sum_{\substack{i \in S: \\ \lvert i/N - z^* \rvert \geqslant\delta}}
    E_i \, v_i^2  \\
  & \geq & \Big (e(z^*)+ \cON \Big )
    \big (1+\cO(\e^{-a N \delta^2}) \big )
          - \gamma \sum_{\substack{i \in S: \\ 
           \lvert i/N - z^* \rvert < \delta}}
           \Big \lvert \frac{i}{N} - z^* \Big \rvert^2 v_i^2 \\
    &&   + \, \cON   + \min_{k \in S} ( E_k  ) 
          \sum_{\substack{i \in S: \\ \lvert i/N - z^* \rvert \geqslant\delta}}
          v_i^2  \\
  & \geq & e(z^*) + \cON.
\end{eqnarray*} 
In the second step, we have  used (A1), normalization 
($\sum_i v_i^2=1$), and ((39)) 
(which also holds for $k=0$, cf.\ Lemma 2 and Cor.\ 2 of \cite{BBBK05}); 
the last step relies on ((39)), ((40)), and (A3).

In the proof of Prop.~4 of the original article,
starting from the second display (p. 97),
we split the sum into
\be{eq:qform}
\begin{split}
\sum_{i,j \in S} v_i F_{ij} v_j = 
& - \sum_{\substack{i \in S: \\ \lvert i/N - z^* \rvert < \delta}} \; 
\sum_{\substack{k \in S-i: \\ \eta(i,k) > 0}}
                       F_{i,i+k} (v_i - v_{i+k})^2 \\
& - \!\!\! \sum_{\substack{i \in S: \\ \lvert i/N - z^* \rvert \geq \delta}} \;
\sum_{\substack{k \in S-i: \\ \eta(i,k) > 0}}
                       F_{i,i+k} (v_i - v_{i+k})^2\,.
\end{split}
\ee
We now note that the display in the middle of p.~97 
implies that, for $ \lvert i/N-z^* \rvert \geq \delta$,
one has $v_i - v_{i+k} \leq c a N \e^{- a N \delta^2}
\eta(i,k)$, where $a$ and $c$ are constants, and
$\eta(i,k) = \cO(1)$ (by the first display on p.~97).
The elements of $\F$ are asymptotically bounded (by (A3)), so the second sum 
in \eqref{eq:qform} is $\cO(N \e^{-\a N \delta^2})$ 
and  plays no
role at the $\cO(1/N)$ level in the remaining calculation.

Let us finally collect the quantities that influence the error term in the
result. These are: the constants in the approximation of $\E$ and $\F$
in (A1); the constant in the decay condition on $f$ in (A2), see Eq.~((45));
the constants in the global
bounds on $\E$ and $\F$ in (A3), as used in
this version of the proof;
and the Hessian of $e$ at $z^*$ (it enters the constant 
$\gamma$). This completes the proof.
\end{proof}
\begin{proof}[of Thm.~\ref{thm:Lambda_e}]
Pick any exposed smoothness point $z\in\sA$, and
let  $\beta=-\nabla e(z)$. Consider the function $\bar e(y):= e(y)+\beta\cdot y$,
$y\in\text{rint\,}\sD$. By hypothesis, $\bar e$ has the unique maximizer $z$. 
Assumptions (A1)--(A3) thus hold for the modified reproduction rates $\bar E_i:= E_i+\beta\cdot i/N$ and the approximating function $\bar e$. (Note that the error
terms do not depend on $\beta$.)  Theorem \ref{thm:lambda_asym} then implies that
$\lambda(\bar E) = \bar e(z)+\cO(1/N)$.
Next we apply Prop.~\ref{prop:Lambda_general} to the type set $S/N$ to infer that
$\lambda(\bar E) = \Lambda(\bar z) + \beta\cdot \bar z$ for the
vector $\bar z=\sp{\bar \alpha,\id/N}=\nabla_{\!\beta\,}
\lambda(E+\beta\cdot\id/N)$, where $\bar\alpha$ is the ancestral
distribution for the reproduction rate $\bar E$. 
(Alternatively, one can invoke Cor.~\ref{cor:Lambda_rev}(a) to
characterize $\bar z$  by the equation $\nabla\Lambda(N\bar z)
=\nabla e(z)$.) 
The comments in paragraph \ref{subsec:approx_lambda}.3 above assert that 
$\bar z =z+\cO(N^{-1/3})$. Hence
\[
\Lambda(\bar z) = \bar e(z) - \beta\cdot \bar z+\cO(1/N)=e(z)+\cO(N^{-1/3})\,.
\]
By the assertion on the error terms in Thm. \ref{thm:lambda_asym} and in
paragraph \ref{subsec:approx_lambda}.3,
the error term here is locally uniform in $z$. 

Next we note that the ($N$-dependent) mapping $\phi:z\to\bar z$ 
from $\sA$ into $\sD$ is a 
homeomorphism. For, $\phi$ is the composition of $-\nabla e$ and 
$\beta\to\nabla_{\!\beta\,}\lambda(E+\beta\cdot\id/N)$.
Now, $\nabla e$ is a diffeomorphism from 
$\sA$ into $T=\text{span}(\sD-\sD)$ because, by assumption, the Hessian of $e$ (restricted to $T$) is nondegenerate everywhere on the convex set $\sA$, so that 
$\nabla e(x)=\nabla e(y)$ only if $x=y$ by the mean value theorem. 
On the other hand,  Corollary~\ref{cor:Lambda_rev}(a) shows that
$\nabla_{\!\beta\,}\lambda(E+\beta\cdot\id/N)$, as a function of $\beta\in T$, has the inverse $y\to-\nabla\Lambda(Ny)$; these gradients are continuous by Corollary 25.5.1 of
\cite{Rock70}.

Now let $\sC\subset\sA$ be compact and $\sC'\subset\sA$, say, a convex polytope
containing $\sC$ in its relative interior. $\phi$ moves the faces of $\sC'$ by at most a
distance of 
$\kappa N^{-1/3}$, 
for some constant $\kappa<\infty$. Hence 
$\phi(\sC')\supset\sC$ for large $N$. For these $N$ we can invert $\phi$ on $\sC$ to get
$\phi^{-1}(y)=y+\cO(N^{-1/3})$ uniformly for all 
$y\in\sC$. Since $|\nabla e|$ is bounded on $\sC$, it follows that
\[
\Lambda(y)=e\big(\phi^{-1}(y)\big)+\cO(N^{-1/3}) = e(y)+\cO(N^{-1/3})
\]
uniformly for all $y\in\sC$.
\end{proof}

\begin{proof}[of Thm.~\ref{thm:basin}]
Take any $z\in\sB$. By a well-known theorem of Carath\'eo\-dory
(Thms.~17.1 and 18.5 of \cite{Rock70}), $z$ is a convex combination
of at most $d+1$ extremal points, that is, there exist points $z_1,\ldots,z_\l\in \text{ex\,}\sB$
and numbers $s_1,\ldots,s_\l\ge 0$ summing to~$1$ such that $\l\le d+1$ and
\[
z=\sum_{k=1}^\l s_k\,z_k\,.
\]
Next we fix some $\eps>0$. By hypothesis, for each $k=1,\ldots,\l$ we can find a point 
$y_k\in\sH\setminus\sB$ and a relatively open convex neighbourhood $\sA_k$ of $y_k$ 
such that  $|y_k-z_k|<\eps$ and $\sA_k$ consists of exposed smoothness points of~$e$.
Theorem~\ref{thm:Lambda_e} thus asserts that 
$\Lambda(y_k)= e(y_k)+\cO(N^{-1/3})$. 
In view of the assumed uniform bounds on the spectrum of the Hessians, 
the error term here is independent of $k$ and the choice of $y_k$. Letting $\eps\to0$, we thus can conclude
that $\Lambda(z_k)= e(z_k)+\cO(N^{-1/3})$, and therefore by concavity
\[
\Lambda(z)\ge\sum_{k=1}^\l s_k\,\Lambda(z_k) 
= \sum_{k=1}^\l s_k\,e(z_k) +\cO(N^{-1/3}) = \hat e(z) +\cO(N^{-1/3})\,.
\]
On the other hand, since the upper estimate on $E_i$ in (A3$'$) also holds for $z=z_k$,
assumptions (A1)--(A3) hold with $\bar E_i = E_i+\beta\cdot i/N$ and $\bar e := e+\beta\cdot\id$ in place
of $E_i$ and $e$, respectively; here $\beta$ is as in \eqref{eq:basin}.
Prop.~\ref{prop:Lambda_general} and Thm.~\ref{thm:lambda_asym} therefore imply
that, for each $k$,  
\[
\Lambda(z) +\beta\cdot z\le \lambda(\bar E)=\bar e(z_k)+ \cO(1/N)\,,
\]
where $\lambda(\bar E)$ stands
for the principal eigenvalue of the matrix $\bar\E+\F$ with reproduction rate $\bar E$.
Taking the average over $k $ we find
\[
\Lambda(z) -\cO(1/N)\le
-\beta\cdot z + \sum_{k=1}^\l s_k\,\bar e(z_k)  = \hat e(z).
\]
The proof is therefore complete.
\end{proof}

\section{Application to the quasispecies model}
\label{sec:quasispec}

\subsection{The model and its large-$N$ asymptotics}

We will now illustrate and apply the results of the preceding Section 
to the coupled counterpart of the parallel  sequence space
model of Subsec.~\ref{subsec:lumping}.
The  coupled sequence space model,  known as the 
\emph{quasispecies model}, was introduced in \cite{Eige71} and  
has, since then, been the subject of
numerous investigations. It assumes that 
mutations occur on the occasion of reproduction events, that is,
they represent replication errors. Let us  assume that mutation is, 
again, independent
across sites and occurs at probabilities $v=\mu/N$ and $w=\nu/N$ from 
$0$ to $1$ and vice versa, where $\mu$ and $\nu$
are positive and independent of $N$. 
This is a slight generalization of the original model \cite{Eige71} with
symmetric mutation, and the factor $1/N$ in the mutation rate is introduced
to obtain a suitable limit\footnote{The factor may come somewhat unexpected, 
but means nothing
but a change of time scale, which  will not alter the long-term asymptotics. 
For a thorough discussion of the related scaling issues, see
\cite{BaWa01}; in the language of that article, we use intensive
scaling here.}. The matrix of mutation probabilities,
$\cP = (\cP_{\s\t})_{\s,\t \in \varSigma^N}$, is then given by 
\be{eq:qs_mut}
  \cP = \bigotimes_{i=1}^N \begin{pmatrix} 1-v & v \\
                                           w & 1-w
                           \end{pmatrix},
\ee
where the tensor product reflects the independence across sites.
The quasi\-species model is complete if we further specify
birth rates $\cB_{\s}$ and death rates $\cD_{\s}$ for all $\s \in \varSigma$. 
When a birth event occurs to a $\s$ individual, it survives
unchanged and produces an offspring
of type $\t$ with probability $\cP_{\s \t}$; at a death
event, a $\s$ individual dies (as in Fig.~\ref{fig:museC} with
$i,j$ replaced by $\s,\t$).

We will assume  that, for all $\s \in \varSigma$,
$\cB_{\sigma}$ and $\cD_{\sigma}$ are invariant under 
permutation of sites.
Since the same holds, by construction, for the mutation probabilities 
\eqref{eq:qs_mut}, we have a situation analogous to (L1) and (L2) 
for the parallel model, and may perform lumping into 
$S:= \{0,1,\ldots,N\}$ by the
mapping $\s \mapsto H(\s) \in S$, where $H(\s)$
is the number of sites occupied by letter 1 (see Sec.~\ref{subsec:lumping}).
The resulting model on $S$ has  birth rates $B_i$, death rates $D_i$, and
mutation probabilities $P_{ij}$, where $\cB_{\s}=B_i$, 
$\cD_{\s}=D_i$, and
\be{eq:qs_lumped}
  \sum_{\t: H(\t)=j} \cP_{\s \t} = P_{ij}
\ee
for any $\s$ with $H(\s)=i$.

In the lumped model, given the current type $i$, the distribution of jumps is
obviously given by the convolution
\be{eq:jump_distr}
  P_{i,i+\bullet} = \Bin_{N-i,v} * \widehat \Bin_{i,w},
\ee
where $\Bin_{n,p}$ denotes the binomial distribution with
parameters $n,p>0$, and
$\widehat \Bin_{n,p}$ its image
under the reflection of $\ZZ$ at the origin;
we further identify $\Bin_{0,p}$ with the point measure
located at $0$. Explicitly, 
\be{eq:qs_explicit}
P_{ij} = (1-v)^{N-i} (1-w)^i 
\sum_{\substack{\l,m \geq 0:\\ \l - m=j-i}} {N-i \choose \l}
{i \choose m} \Big ( \frac{v}{1-v} \Big )^{\l} 
\Big ( \frac{w}{1-w} \Big )^{m}.
\ee
The Markov chain so defined is reversible with respect to
$\ph = (\ph_i)_{i \in S} = \Bin_{N,v/(v+w)}$; this is most easily seen by noting
that $\cP$ (on sequence space) is reversible with respect to the
Bernoulli measure on $\{0,1\}^N$ with parameter $v/(v+w)$.

The  lumped Markov branching process  has first-moment generator $\A$ with
elements $A_{ij} = B_i P_{ij} - D_i$ (cf.\ Eq.~\eqref{eq:A_coupled}),
and has been much studied,  see \cite{EMcCS89} for a review of early work,
and \cite{Kamp03} for a  review of recent theoretical developments,
and their connection to experimental results on virus evolution.
In particular, the error
thresholds displayed by this model have attracted a lot of attention.

The function $e(z)$ that would simplify the model's analysis does not
seem to have appeared so far; it is far less obvious than
its parallel counterpart \eqref{eq:e_parallel}, and will be established 
in what follows. We start by
decomposing $\A$ of \eqref{eq:A_coupled}
into a Markov generator $\U$ and a diagonal matrix $\R$, which 
gives $U_{ij}=B_iP_{ij}$ for $i \neq j$, 
$U_{ii}= - B_i (1-P_{ii})$, and 
$R_i= \sum_{j \in S} A_{ij} = B_i - D_i$. Since
$\P=(P_{ij})_{i,j \in S}$ is reversible, $\U$ is also reversible;
its reversible distribution
$\rho$ is given by $\rho_i=c \varphi_i / B_i$ for a normalizing
constant $c>0$.
The elements of the symmetrized matrices $\E$ and $\F$ of \eqref{eq:E}
and \eqref{eq:F} therefore emerge as
\be{eq:F_ij_coupled}
 F_{ij}=\sqrt{B_i P_{ij} P_{ji} B_j}\quad \text{for} \;  i \neq j,
\ee
\be{eq:F_ii_coupled}
F_{ii}=-\sum_{j \in S: j \neq i} F_{ij},
\ee 
and 
\be{eq:E_i_coupled}
E_i = - D_i + \sum_{j \in S} \sqrt{B_i P_{ij} P_{ji} B_j}.
\ee

After these preparations, let us identify conditions under which 
Thms.~\ref{thm:lambda_asym}, \ref{thm:Lambda_e}, and \ref{thm:basin}
are applicable. We will consider the  approximation of
the birth and death rates as given;
we will then show that the Poisson
approximation to the distribution $P_{i,i+\bullet}$, namely $p_{\bullet}(i/N)
=  \Poi_{\mu (N-i)/N} * \widehat \Poi_{\nu i/N}$,
will also lead to
the `right' approximation to the matrix elements 
\eqref{eq:F_ij_coupled}--\eqref{eq:E_i_coupled}. In line with previous
notation, $\Poi_{\lambda}$
is the Poisson distribution with parameter ${\lambda>0}$,
$\widehat \Poi_{\lambda}$ its  reflected version, and $\Poi_0$
is identified with the point measure at $0$. 
This will give us the following result.
\begin{theorem}
\label{thm:quasispec}
Consider the lumped quasispecies model, with first-moment
generator $\A$ of \eqref{eq:A_coupled} on $S=S(N) := \{0,1,\ldots,N\}$;
birth rates $B_i \geq 0$, 
death rates $D_i \geq 0$, and mutation probabilities $P_{ij}$ as in
\eqref{eq:qs_explicit}. Assume that 
\[
B_i=b \biN+\cON \quad\text{ and }\quad D_i=d\biN+\cON,
\]
where $b$ and $d$ are $C^2$  functions on $\sD := [0,1]$, 
$b$ is strictly positive, and the
constants in the $\cO(1/N)$ bounds are uniform for all $i \in S$.
For $z\in\sD$ let 
\be{eq:g}
  g(z) := \big( \sqrt{\mu (1-z)} - \sqrt{\nu z}  \big)^2
\ee
and
\be{eq:e_quasispec}
e(z) := b(z) \,\e^{- g(z)} - d(z)\,.
\ee
Assume further that $e''$ has only finitely many zeroes. 
It then follows that
\[
  \lambda  = \max_{z\in\sD}e(z) + \cON \quad\text{and}\quad \Lambda(z) = e(z)+\cO(N^{-1/3})
\]
locally uniformly for $z\in\ ]0,1[$.
\end{theorem}

Postponing the proof for a moment, let us first look at an example. 

\begin{figure}[h]
\begin{center}
\psfrag{A}{\small$z$}
\psfrag{B}{\small$r(z)$}
\psfrag{C}{\small$z$}
\psfrag{D}{\small$e(z)$}
\psfrag{E}{\small$\hat e(z)$}
\epsfig{file=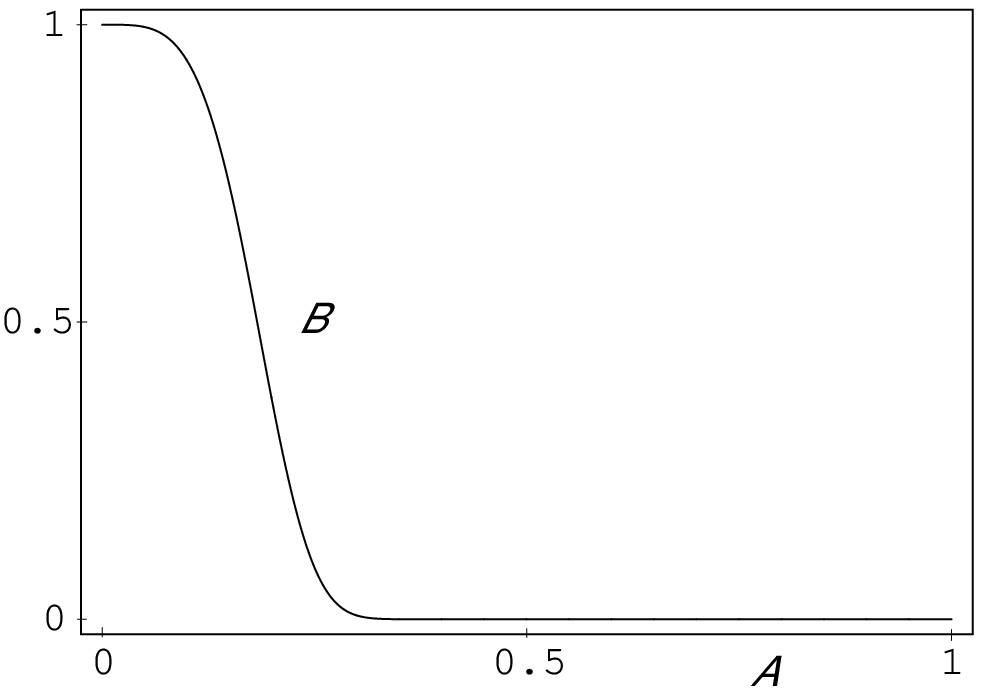, height=3.7cm}\hskip3mm 
\epsfig{file=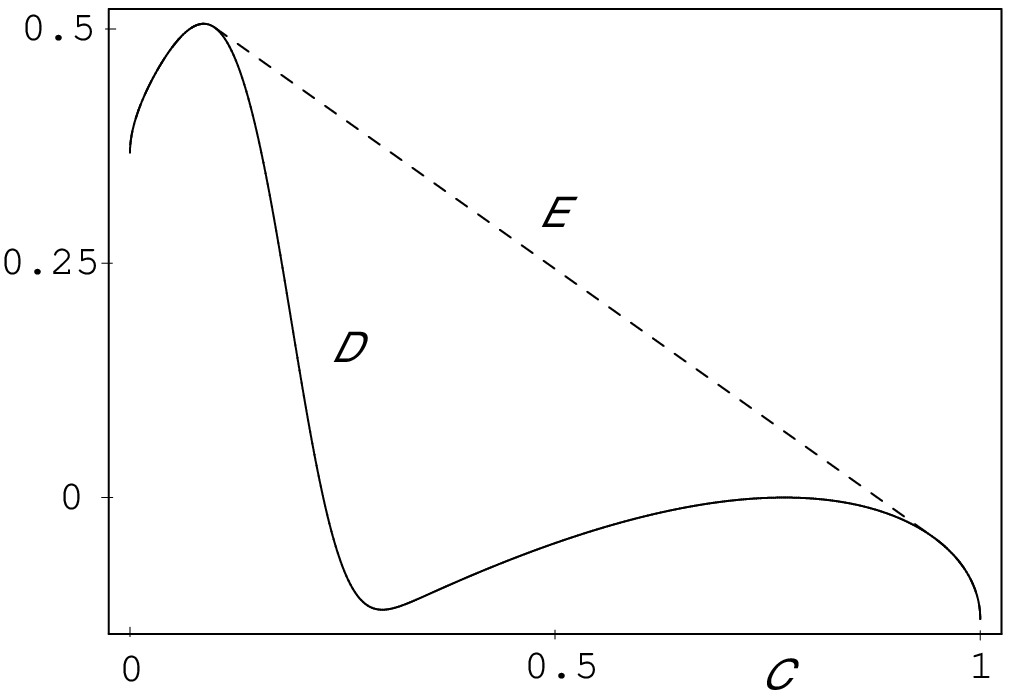, height=3.7cm}
\end{center}
\caption{\label{fig:quasispec} The quasispecies example \eqref{eq:quasi_ex},
  with $\gamma=5.$, and $\mu=1$, $\nu=0.3$.
  Left: The fitness function, $r=b-d$.  Right: The
  function $e$ (solid line) and its concave envelope $\hat e$
(dashed), where it
  deviates from $e$.}
\end{figure}

\smallskip
\emph{\thesubsection.1 An example.}
For the purpose of illustration, let us consider the quasispecies model
with a `smoothed' version of {\em truncation selection} (where a
gene tolerates a certain number of mutations and then deteriorates
rapidly). Let the birth and death rate functions be given by
\be{eq:quasi_ex}
  b(z) := \frac{1+r(z)}{2}, \quad d(z) := \frac{1-r(z)}{2},
  \quad \text{where } r(z) := \e^{-(\gamma z)^4}
\ee
(i.e., we assume a mixture of fecundity  and viability selection). 
Fig.~\ref{fig:quasispec} shows
the fitness function, and the function $e$ together with its
concave envelope. 

\emph{\thesubsection.2 Connection to the parallel model.}
The quasispecies  model is  closely related to the lumped  {\em parallel}
sequence space model with birth rates $B_i$, death rates $D_i$,
and mutation rates $U_{i,i+1}= \mu (1-i/N)$,  $U_{i,i-1}= \nu\, i/N$
and $U_{ij}=0$ for $\lvert j-i \rvert > 1$
(where $\mu$ and $\nu$ are now mutation rates per site
rather than probabilities). In fact, the latter may be considered as 
the former's weak-selection weak-mutation
limit (cf.\ \cite[Ch.II.1.2]{Buer00}, 
and \cite{Hofb85}). It leads to the simpler expression
\be{eq:e_ss_parallel}
   e(z) = b(z) - d(z)  - g(z),     
\ee
cf.\ \eqref{eq:e_parallel}, and \cite{BBBK05,HRWB02}. 
Indeed, this function is easily
identified as the weak-selection weak-mutation limit of 
\eqref{eq:e_quasispec} by 
replacing $b$ by $1+\delta b$, $d$ by $1+\delta d$, $\mu$ by $\mu \delta$,
$\nu$ by $\nu \delta$, and $e$ by $e/\delta$; the last replacement means
that time is measured in units of $\delta$. $e(z)$ of \eqref{eq:e_ss_parallel} then emerges from \eqref{eq:e_quasispec} in the limit $\delta\to0$.

\subsection{Proof}

The proof of Thm. \ref{thm:quasispec} consists in verifying the assumptions of
Thms.~\ref{thm:lambda_asym}, \ref{thm:Lambda_e} and  \ref{thm:basin} for the matrices 
$\E$ and $\F$ in \eqref{eq:F_ij_coupled}--\eqref{eq:E_i_coupled}.
The main difficulty will be to establish the $\cO(1/N)$
approximation as required in (A1).
Besides the approximating function $e$ in \eqref{eq:e_quasispec} for $\E$,
the approximating functions $f_k$ for $\F$ will be given by
\be{eq:f_k}
   f_0(z) := b(z) \big(p_0(z)-\e^{-g(z)} \big), \quad
   f_k(z) := b(z) \sqrt{p_k(z) p_{-k}(z)} \quad (0\ne k \in \ZZ), 
\ee
where
\be{eq:p_k}
  p_k(z):=  (\Poi_{\mu (1-z)} \ast \widehat \Poi_{\nu z}  ) (k),
  \; k \in \ZZ.
\ee
These functions are quite natural, as they are obtained by
replacing the binomial distributions
at hand by their Poisson approximations. Nevertheless, the required approximation
result is not at all automatic: Although $\Bin_{\mu,(N-i)/N}$ and $\widehat \Bin_{\nu,i/N}$
deviate from $\Poi_{\mu (N-i)/N}$ and $\widehat \Poi_{\nu i/N}$, respectively,
by $\cO(1/N)$ in variational distance \cite[Section II.5]{Lind92}, 
and this carries over to the  convolution,  it remains to be shown
that the corresponding \emph{symmetrized} quantities share this
property.
The key to this task is the fact that the Poisson distributions are particularly well-suited for a geometric symmetrization as in \eqref{eq:F_ij_coupled}. This is the content of the
following lemma.

\begin{lemma}\label{lem:Poiss_symm}
For $a,b\geq 0$, let
$
p_{\bullet}(a,b)=(p_k(a,b))_{k \in \ZZ}:= \Poi_a \ast \widehat \Poi_b
$
be the convolution of the parameter-$a$ Poisson distribution with the reflected 
parameter-$b$ Poisson distribution. Then
\[
\sqrt{p_k(a,b)p_{-k}(a,b)} = \e^{-(\sqrt{a}-\sqrt{b})^2}\;
p_k(\sqrt{ab},\sqrt{ab})
\]
for all $k\in \ZZ$.
\end{lemma}
\begin{proof}
Since $p_k(a,0)=0$ for $k < 0$, and $p_k(0,b)=0$ for $k>0$, the 
conclusion is immediate if either $a$ or $b$ vanishes. For $a,b>0$, 
the explicit formula
\be{eq:p_kab} 
  p_k(a,b) = \e^{ -a-b}
   \sum_{\substack{\l,m\geq 0: \\ \l-m=k}} \frac{ a^{\l} b^{m}}{\l! m!}
\ee
readily implies that $  p_{-k}(a,b) = p_k(a,b) ( b/a )^k$, whence 
\be{eq:sqrt_p_kp_-k}
\sqrt{p_k(a,b)p_{-k}(a,b)}=p_k(a,b) (b/a)^{k/2}
\ee
for all $k\in\ZZ$. Inserting 
\eqref{eq:p_kab} into the last term
and comparing the result with the similar expression for 
$p_k(\sqrt{ab},\sqrt{ab})$ we obtain the conclusion of the lemma.
\end{proof}

\medskip
We will be particularly interested in the Poisson approximation to 
the right-hand side of \eqref{eq:jump_distr}, viz.
\be{eq:p_kz}
p_k(z):= p_k (\mu(1-z),\nu z)\,,
\ee
where $z:=i/N$. 
Lemma \ref{lem:Poiss_symm} then implies that
\be{eq:Poiss_symm_sum}
\sum_{k\in\ZZ}\sqrt{p_k(z)p_{-k}(z)} = \e^{-g(z)}\,,
\ee
thereby explaining the origin of the function $g$ defined in \eqref{eq:g}.
We will also need the following tail estimate.

\begin{lemma}\label{lem:Poiss_tail}
For all $a,b\geq 0$,
\[
\sum_{k\in\ZZ:\,|k|\ge\sqrt N}  \sqrt{p_k(a,b)p_{-k}(a,b)} \leq \frac {2 \sqrt{ab}}N\,.
\]
\end{lemma}
\begin{proof}
If $a=0$ or $b=0$,  $p_k(a,b)p_{-k}(a,b)=0$ except for $k=0$,
so that the assertion is trivial. For $a,b>0$ we can write,
using Lemma \ref{lem:Poiss_symm} and Markov's inequality:
\[
N\sum_{k\in\ZZ:\,|k|\ge\sqrt N}  \sqrt{p_k(a,b)p_{-k}(a,b)}
\le \sum_{k\in\ZZ} k^2\,p_k(\sqrt{ab},\sqrt{ab})\,.
\]
By symmetry, the last sum is the variance of 
$\Poi_{\sqrt{ab}} \ast \widehat\Poi_{\sqrt{ab}}$
and thus equal to $2 \sqrt{ab}$.
\end{proof}

\medskip
We need a similar tail estimate for the geometric symmetrization of the matrix 
$\P$ defined in \eqref{eq:qs_explicit}. Note that $\P$ depends on $N$.

\begin{lemma}\label{lem:Bin_tail}
For all $i\in S$,
\[
\sum_{j\in S:\,|j-i|\ge\sqrt N}  \sqrt{P_{ij}P_{ji}} \leq \frac CN
\]
for a constant $C$ depending on $\mu$ and $\nu$ but not on $N$.
\end{lemma}
\begin{proof}
We use the obvious inequality
\be{eq:Bin_leq_Poiss}
\Bin_{n,p}(\l) = (1-p)^n {n\choose \l}\Big(\frac{p}{1-p}\Big)^\l
\le \e^{a-np}\, \Poi_{a}(\l)
\ee
which holds whenever $0\le \l\le n$, $0<p<1$, and $a \geq np/(1-p)$. This implies
that $P_{i,i+k}\le \e^{a+b}\, p_k(a,b)$ with $a:= \mu/(1-v)$, $b:= \nu/(1-w)$,
uniformly in $i\in S$. Hence
\be{eq:universal_sqrt_bound}
 \sqrt{P_{i,i+k}P_{i+k,i}} \le \e^{a+b}\,  \sqrt{p_k(a,b)p_{-k}(a,b)}\,,
\ee
and the result follows from Lemma \ref{lem:Poiss_tail}.
\end{proof}

The crucial step is the following Poisson approximation
of the geometric symmetrization of $\P$. 
 
\begin{proposition}\label{prop:approx_sqrt}
With the abbreviation $p_k(z):= p_k (\mu(1-z),\nu z)$, we have
  \be{eq:sum_sqrt}
  \sum_{j \in S}\Big |\sqrt{P_{ij} P_{ji}} 
   -\sqrt{p_{j-i} ( i/N) p_{i-j}(i/N)}\Big|
   = \cON  
\ee 
uniformly in $i$ as long as $i/N$ is bounded away from $0$ and $1$.
\end{proposition}
\begin{proof}
Consider an arbitrary $i\in S$ 
and suppose $z:=i/N$ is bounded away from $0$ and $1$. 
(We will generally suppress the 
$i$-dependence of all abbreviations to be introduced below.) 
The main difficulty of the proof comes from
the fact that the parameters of the probabilities $P_{ji}$ depend on the variable
$j$ rather than $i$. Fortunately, $\P$ is reversible, and
Lemmas \ref{lem:Poiss_tail}
and \ref{lem:Bin_tail} allow us to confine ourselves to the $j$'s with 
$|j-i|\le\sqrt{N}$.
We proceed by a comparison of upper and lower bounds on
\[
r_k:=\sqrt{P_{i,i+k}P_{i+k,i}}\,,\quad k\in S-i \,.
\]

\emph{Step 1: A lower estimate. }
Since $\P$ is reversible w.r.t.\ $\Bin_{N,v/(v+w)}$, we have
\[
  r_k   =  P_{i,i+k} \sqrt{\frac{(i+k)! (N-i-k)!}{i! (N-i)!}} 
\Big ( \sqrt{\frac{w}{v}} \Big )^{k} 
\]
for all $k\in S-i$. Since also $m!/n! \geq n^{m-n}$ for
all $m,n$, it follows that
\be{eq:lobound}
r_k\ge \underline{r}_k := P_{i,i+k}\, s^{k}  
\ee
for all $k$, where 
$s:= (iw/(N-i)v)^{1/2}=(z\nu/(1-z)\mu)^{1/2}$. 

We now take the sum over $k$. Using 
\eqref{eq:qs_explicit} and the binomial formula 
we can write
\be{eq:sumlobound}
\begin{split}
  \sum_{k\in S-i} \underline{r}_k 
 &=  ( 1 - v + vs)^{N-i}( 1 - w + w/s )^{i} \\
  \ge\; & \exp\big(-\mu(1-s)(1-z) \big)
     \Big (1- \frac{\mu^2 (1-s)^2(1-z)^2}{N-i} \Big ) \\
  &\times 
      \exp\big(-\nu(1-1/s)z\,\big) 
      \Big (1 -  \frac{\nu^2(1-1/s)^2z^2}{i} \Big ) \\
  =\;& \e^{ - g(z)}  + \cON \,.
\end{split}
\ee
The inequality follows from the fact that
\be{eq:mitr}
 \Big ( 1 - \frac{x}{n} \Big )^n \geq \e^{-x} \Big(1 -  \frac{x^2}{n}\Big)
\ee
for any $n \geq 1$ and $\lvert x \rvert \leq n$; see
\cite[3.6.2., p.~266]{Mitr70}.
In the last step we used that $z$ is bounded away from $0$ and $1$.

\emph{Step 2. An upper bound. } Arguing as for 
\eqref{eq:lobound} we find
\[
r_k  \le P_{i,i+k} \Big ( \frac{(i+k)w}{(N-i-k)v} \Big )^{k/2} 
= P_{i,i+k}\,s^k \, \big ( \psi(k/N) \big )^{k}\,,
\]
where $\psi(x):= (1+ x/z )^{1/2}(1- x/ (1-z) )^{-1/2}$.
Next, for each $k$ an expansion of $\psi$  gives
\[
\big ( \psi ({k}/N ) \big )^{k} =
 1+  \frac{k^2}{N} \; 
\big ( \psi (\vartheta_k ) \big )^{k-1} \psi'(\vartheta_k)
\]
for some $\vartheta_k$ between $0$ and $k/N$. As long as $z$ is bounded 
away from $0$ and $1$  and $|k|\le\sqrt{N}$,
the $\psi$- and $\psi'$-expressions on the right-hand side
are bounded from above, so that
\[
\big ( \psi ({k}/N ) \big )^{k}  \le  1+ \frac{k^2\,c^k}{N} 
\]
for some $c<\infty$. Also, using \eqref{eq:Bin_leq_Poiss} we find
\be{eq:ab_bound}
P_{i,i+k} \le \e^{va+wb}\, p_{k} (a,b) \quad\text{ with }
\quad a:=\frac{(N-i)v}{1-v},
\; b:=\frac{iw}{1-w}\,.
\ee
Collecting all estimates we arrive at the upper bound
\be{eq:upbound}
r_k \le \bar{r}_k := \e^{va+wb}\, p_k (a,b) s^k  
\,\Big(1+ \frac{k^2\,c^k}{N}\Big) \,.
\ee

Next, a summation over $k$ gives
\[
\sum_{|k|\le\sqrt N} \bar{r}_k\le
 \e^{va+wb}\,\Big(\sum_{k \in \ZZ} p_k(a,b) s^{k} +\frac{K}{N} \Big)
\]
with 
$K:=\sum_{k \in \ZZ}k^2\,(cs)^k \, p_{k} (a,b)$.
To deal with the terms on the right-hand side we note first that
$ p_k(a,b)\, s^{k}=\e^{as+b/s-a-b} p_k(as,b/s)$
by formula \eqref{eq:p_kab}. 
Hence
\[
\sum_{k \in \ZZ} p_k(a,b)\, s^{k} = \exp(as+b/s-a-b) 
=\exp\big(-g(z) \big) +\cON
\]
because $a\approx(1-z)\mu$ and $b\approx z\nu$ up to error terms of
order $1/N$. Likewise,
\[
K = \exp\big(acs+b/(cs)-a-b\big)\sum_{k \in \ZZ}k^2\,p_k\big(acs,b/(cs)\big)
\]
is bounded in $N$ since so are $a,b$. Since also $va+wb=\cO(1/N)$, we 
finally arrive at the estimate
\be{eq:sumupbound}
\sum_{|k|\le\sqrt N} \bar{r}_k \le \e^{-g(z)}  +\cON\,.
\ee

\emph{Step 3: Conclusion. }Consider now $q_k:= \sqrt{p_k (z) p_{-k}(z)}$.
By \eqref{eq:sqrt_p_kp_-k}, $q_k = p_k(z) s^k$. It is also
immediate that $p_k(z) \le \e^{va+wb}\, p_k (a,b)$ with $a,b$ as in 
\eqref{eq:ab_bound}. Hence $ q_k \le\bar{r}_{k}$ for all $k$. 
Combining this with Lemmas \ref{lem:Poiss_tail} and  \ref{lem:Bin_tail}
and the bounds \eqref{eq:lobound} and \eqref{eq:upbound} we find
  \[
 \begin{split}
 \sum_{k \in S-i}  \lvert r_k-q_k \rvert
&= \sum_{k \in S-i} \big(2 \max(r_k,q_k) - r_k-q_k\big) \\
  &\leq  \cON+  2\sum_{|k|\le\sqrt N} \bar{r}_k-\sum_{k \in S-i}\underline{r}_k
-\sum_{|k|\le\sqrt N}  q_k\,.
 \end{split} 
 \]
Now,  \eqref{eq:sumlobound} and \eqref{eq:sumupbound} show that the first sum 
exceeds the second only by a term of order $1/N$, and \eqref{eq:sumupbound} 
together with \eqref{eq:Poiss_symm_sum} and Lemma \ref{lem:Poiss_tail}
imply that the first sum exceeds the third by at most a term of order $1/N$.  
This completes the proof.
\end{proof}

Besides the preceding key approximation in the interior of $\sD$,
we also need a uniform bound which will be used close to the boundary
of $\sD$. Here, an error bound of order $1/\sqrt{N}$ is sufficient.

\begin{proposition}\label{prop:upper_bound}
For all $i \in S$, 
\[
\sum_{j\in S}  \sqrt{P_{ij}P_{ji}} \leq \e^{-g(i/N)} + \, \cOsN,
\]
where the $\cO(1/\sqrt{N})$ term depends on $\mu$ and $\nu$ 
but not on $i$.
\end{proposition}

\begin{proof}
In view of Lemma \ref{lem:Bin_tail}, we only need to estimate the sum over all
$j= i+k \in S$ with $|k| \leq \sqrt{N}$. For these $j$ we find,
using the Poisson bound~\eqref{eq:Bin_leq_Poiss} and  writing again $z=i/N$,
\[
 \sqrt{P_{i,i+k}P_{i+k,i}} \le 
 \e^{\tilde a + \tilde b}\,  \sqrt{p_k(a,b)p_{-k}(a,b)}.
\]
Here 
$ a  = \mu(1-z + {1}/{\sqrt{N}})/(1-v) $, 
$ b  = \nu(z + {1}/{\sqrt{N}})/(1-w) $,
and
\[
\begin{split}
\tilde a &= a - \mu(1-z - 1/\sqrt{N})=2\mu /\sqrt{N} + \cO(1/N) \,,\\
\tilde b &= b - \nu(z - 1/\sqrt{N})= 2\nu /\sqrt{N} + \cO(1/N)\,,
\end{split} 
\]
so that $\e^{\tilde a + \tilde b} = 1 + \cO(1/\sqrt N)$;
the error terms do not depend on $i$. The claim thus follows
from Lemmas \ref{lem:Bin_tail}, \ref{lem:Poiss_tail}, \ref{lem:Poiss_symm}
and the fact that $(\sqrt{a}-\sqrt{b})^2=g(z)+\cO(1/\sqrt{N})$.
\end{proof}

After these preparations we are now ready to read off the
approximating functions $e$ and $f_k$ for the lumped quasispecies model,
that is, we can  proceed to the

\begin{proof}[of Thm.~\ref{thm:quasispec}]
The main point of the proof is to establish condition (A1) for any compact interval $\sA\subset\,]0,1[$.
The (asymptotic) boundedness of the $B_j$'s and Lemma \ref{lem:Bin_tail} 
imply that, for each $i\in S$,
\be{eq:F_approx_1}
\sum_{k \in S-i}\sqrt{B_i B_{i+k} P_{i,i+k} P_{i+k,i}} = \cON +
\sum_{ \lvert k \rvert\leq \sqrt N }\sqrt{B_i B_{i+k} P_{i,i+k} P_{i+k,i}}\,.
\ee
The asymptotics of the $B_j$'s implies further that the sum
on the right-hand side is equal to
\be{eq:F_approx_2}
\Big(1+\cON\Big) \;b(z)\sum_{ \lvert k \rvert\leq \sqrt N }
\exp \big (\b(z+k/N)-\b(z)\big)\sqrt{P_{i,i+k} P_{i+k,i}}\,,
\ee
where  $z:=i/N$ and $\b(x):=(\log b(x))/2$. By hypothesis, $\b\in C^2_b([0,1])$. Hence 
$\b(z+k/N)-\b(z)=\b'(z)k/N + \cO(1/N)$ for $|k|\leq \sqrt N $, so that the last expression
takes the form
\be{eq:F_approx_3}
\Big(1+\cON\Big) \;b(z)\sum_{ \lvert k \rvert\leq \sqrt N }
\e^{k\d/N}\sqrt{P_{i,i+k} P_{i+k,i}}\,,
\ee
where $\d:= \b'(z)$. Next we can omit the exponential $\e^{k\d/N}$, making an error
of order $1/N$ only. Indeed, using inequality \eqref{eq:universal_sqrt_bound}
together with Lemma \ref{lem:Poiss_symm} and setting  $a:= \mu/(1-v)$, 
$b:= \nu/(1-w)$ we obtain
\be{eq:F_approx_4}
\begin{split}
\sum_{ k\in S-i}\lvert \e^{k\d/N}-1\rvert&\sqrt{P_{i,i+k} P_{i+k,i}}
\\
&\le \e^{a+b}\sum_{ k\in\ZZ}\big( \e^{|k| |\d|/N}-1\big)\,
p_k(\sqrt{ab},\sqrt{ab})
\\
&\le \e^{a+b}\Big(\exp\big[2 \sqrt{ab}\,(\e^{|\d|/N}-1)\big]-1\Big) =\cON\,.
\end{split}
\ee
The second inequality is obtained by taking formula \eqref{eq:p_kab}
for $p_k(\sqrt{ab},\sqrt{ab})$, 
using $\e^{|k| |\d| /N}\le \e^{\l |\d| /N}\e^{m|\d|/N}$, and summing up.
If $z=i/N$ is bounded away 
from $0$ and $1$, we can finally apply Prop.\ \ref{prop:approx_sqrt}, 
Lemma \ref{lem:Poiss_tail} 
and the identity \eqref{eq:Poiss_symm_sum} to obtain
\be{eq:F_approx_5}
\begin{split}
&\sum_{k \in S-i}\sqrt{B_i B_{i+k} P_{i,i+k} P_{i+k,i}} \\ 
&\qquad= \cON +
\Big(1+\cON\Big) \;b(z)\sum_{ \lvert k \rvert\leq \sqrt N }
\sqrt{p_k (z) p_{-k}(z)}\\
&\qquad=b(z)\e^{-g(z)} + \cON \,.
\end{split}
\ee
Taking this together with the assumed approximation of the $D_i$,
we arrive at the approximation (A1) of $\E$, and the diagonal elements
of $\F$, by the functions $e$ and $f_k$ defined in \eqref{eq:e_quasispec} 
and \eqref{eq:f_k}. 
But the approximation of the nondiagonal elements of $\F$
is also guaranteed, since the estimates leading to \eqref{eq:F_approx_5}
all hold term by term. This completes the proof of (A1). 

Next, condition (A2) follows  directly from Lemma
\ref{lem:Poiss_symm} and the fact that
Poisson distributions and their convolutions have a finite third moment.
We note further that the upper bound on $F_{ij}$ and 
the lower bound on $E_i$ in (A3) and (A3$'$) are obvious. 

Before turning to the upper bound on $E_i$ let us discuss the particular context of 
Theorems~\ref{thm:lambda_asym}, \ref{thm:Lambda_e} and  \ref{thm:basin}. 
We observe that the function $g$ is continuous on 
$[0,1]$ and smooth on $]0,1[$ with $g'(0)=-g'(1)=-\infty$, while 
$b>0$ and $d$ are $C^2$ functions on $[0,1]$. This entails that
$e$ is $C^2$ on $]0,1[$ and attains its absolute maximum at a point $z^*\in\:]0,1[$;
in particular, $z^*$ is contained in an interval $\sA$ satisfying (A1) and (A2),
as is required for Thm.~\ref{thm:lambda_asym}. 
In addition, $e''$ is negative in a neigbourhood of $0$ and $1$, and
has only finitely many zeroes by assumption. This implies that each basin of
$e$ is determined by smooth hills, as is necessary for applying Thm.~\ref{thm:basin}.

Now let $\sA'$ be any set of exposed smoothness points of $e$ which is
bounded away from $0$ and $1$. If $\delta >0$ is sufficiently small, the set 
$\sA_\d'$ of all $y$ satisfying  $e(y)\ge t_z(y)-\delta$ for all $z\in\sA'$  is
still bounded away from $0$ and $1$. For all $i$ with $i/N\in\sA_\d'$, 
the upper bound on $E_i$ in (A3$'$) follows directly from \eqref{eq:F_approx_5}. 
For all other $i$'s, this bound follows from 
\eqref{eq:F_approx_1}--\eqref{eq:F_approx_4} as soon as $N$ is so large that the
$\cO(1/\sqrt{N})$-term in Prop.~\ref{prop:upper_bound} is less than $\d/\max b$. 
This completes the proof of (A3$'$)
under the conditions of Thms.~\ref{thm:Lambda_e} and  \ref{thm:basin}.
Since $]0,1[$ splits into finitely many intervals forming basins and smooth hills of $e$,
the stated approximation result for $\Lambda$ follows.
Finally, as $z^*$ is also an exposed point, the choice $\sA'=\{z^*\}$
gives us the upper bound on $E_i$ in (A3). Thm.~\ref{thm:lambda_asym}
can therefore be applied, proving the approximation of $\lambda$ as stated.
\end{proof}

\begin{acknowledgement}
It is our pleasure to thank Joachim Hermisson, who first conjectured
the form of the function $e(z)$ in the lumped quasispecies model;
Michael Baake for helpful discussions on Thm.~\ref{thm:lambda_asym};
and Jesse Taylor for critically reading the manuscript.
\end{acknowledgement}


\begin{thebibliography}{10}

\bibitem{Akin79}
E.~Akin:
\newblock {The Geometry of Population Genetics}. 
\newblock Springer, Berlin, 1979.

\bibitem{AtNe72}
K.~B. Athreya, P.~E. Ney:
\newblock {Branching Processes}.
\newblock Springer, New York, 1972.

\bibitem{BBBK05}
E.~Baake, M.~Baake, A.~Bovier, M.~Klein:
\newblock An asymptotic maximum principle for essentially nonlinear evolution
  models.
\newblock {\em J.\ Math.\ Biol.} {\bf 50}, 83--114 (2005);
\newblock ArXiv:q-bio.PE/0311020

\bibitem{BaWa01}
E.~Baake, H.~Wagner:
\newblock Mutation-selection models solved exactly with methods from
  statistical mechanics.
\newblock {Genet. Res.} {\bf 78} 93--117 (2001)

\bibitem{Buer00}
R.~B{\"u}rger:
\newblock {The Mathematical Theory of Selection, Recombination, and
  Mutation}.
\newblock Wiley, Chichester, 2000

\bibitem{CrKi70}
J.~F. Crow,  M.~Kimura:
\newblock {An Introduction to Population Genetics Theory}.
\newblock Harper \& Row, New York, 1970

\bibitem{DeZe98}
A.~Dembo, O.~Zeitouni:
\newblock {Large Deviations: Techniques and Applications}.
\newblock Springer, New York, 1998


\bibitem{Edwa02}
A.~Edwards:
\newblock The fundamental theorem of natural selection.
\newblock {\em Theor.~Pop.~Biol.} {\bf 61}, 335--337 (2002)

\bibitem{Eige71}
M.~Eigen:
\newblock Selforganization of matter and the evolution of biological
  macromolecules.
\newblock {\em Naturwiss.} {\bf 58}, 465--523 (1971)

\bibitem{EMcCS89}
M.~Eigen, J.~McCaskill, P.~Schuster:
\newblock The molecular quasi-species.
\newblock {\em Adv.\ Chem.\ Phys.} {\bf 75}, 149--263 (1989)

\bibitem{Ewen04}
W.~Ewens:
\newblock {Mathematical Population Genetics}.
\newblock Springer, Berlin, 2nd ed., 2004

\bibitem{EwGr05}
W.~Ewens, G.~Grant:
\newblock {Statistical Methods in Bioinformatics}.
\newblock Springer, New York, 2nd ed., 2005

\bibitem{Gars05}
T.~Garske:
\newblock Error thresholds in a mutation-selection model with {H}opfield-type
  fitness.
\newblock {B}ull.\ {M}ath.\ {B}iol., in press, arXiv:q-bio.PE/0505056;
doi 10.1007/s11538-006-9072-1.

\bibitem{GaGr04}
T.~Garske, U.~Grimm:
\newblock A maximum principle for the mutation-selection equilibrium of
  nucleotide sequences.
\newblock {\em Bull.\ Math.\ Biol.} {\bf 66}, 397--421 (2004);
\newblock arXiv:physics/0303053v2

\bibitem{GeBa03}
H.-O. Georgii, E.~Baake:
\newblock Multitype branching processes: the ancestral types of typical
  individuals.
\newblock {\em Adv.\ Appl.\ Prob.} {\bf 35}, 1090--1110 (2003);
\newblock arXiv:math.PR/0302049

\bibitem{GeHw02}
U.~Gerland, T.~Hwa:
\newblock On the selection and evolution of regulatory {DNA} motifs.
\newblock {J. Mol. Evol.} {\bf 55} 386--400 (2002)

\bibitem{HSW05}
J.~Hein, M.~Schierup, C.~Wiuf:
\newblock {Gene genealogies, variation and evolution : a primer in coalescent 
theory}.
\newblock Oxford University Press, Oxford, 2005

\bibitem{HRWB02}
J.~Hermisson, O.~Redner, H.~Wagner, E.~Baake:
\newblock Mutation-selection balance: Ancestry, load, and maximum principle.
\newblock {\em Theor.\ Pop.\ Biol.} {\bf 62}, 9--46 (2002);
\newblock arXiv:cond-mat/0202432

\bibitem{Hofb85}
J.~Hofbauer:
\newblock The selection mutation equation.
\newblock {\em J. Math. Biol.} {\bf 23}, 41--53 (1985)

\bibitem{deHo00}
F.~den Hollander:
\newblock {Large Deviations}, 
\newblock AMS, Providence, RI, 2000

\bibitem{JaNe84}
P.~Jagers, O.~Nerman:
\newblock The stable doubly infinite pedigree process of supercritical
branching populations.
\newblock {\em Zeitschrift f\"ur Wahrscheinlichkeitstheorie und
verwandte Gebiete} {\bf 65}, 445-460 (1984)

\bibitem{Jage89}
P.~Jagers:
\newblock General branching processes as {M}arkov fields.
\newblock {\em Stoch.~Proc.~Appl.} {\bf 32}, 183--242 (1989)

\bibitem{Jage92}
P.~Jagers:
\newblock Stabilities and instabilities in population dynamics.
\newblock {\em J.~Appl.~Prob.} {\bf 29}, 770--780 (1992)

\bibitem{Kamp03}
C.~Kamp:
\newblock A quasispecies approach to viral evolution in the context of an
  adaptive immune system.
\newblock {\em Microbes and Infection} {\bf 5}, 1397--1405 (2003)

\bibitem{KaTa75}
K.~S. Karlin, H.~M. Taylor:
\newblock {A first course in stochastic processes}.
\newblock Academic Press, San Diego, CA, 2nd edition, 1975

\bibitem{KeSn81}
J.~G. Kemeny, J.~L. Snell:
\newblock {Finite {M}arkov Chains}.
\newblock Springer, New York, 1981

\bibitem{KeSt66}
H.~Kesten, B.~P. Stigum:
\newblock A limit theorem for multidimensional {G}alton-{W}atson processes.
\newblock {\em Ann.\ Math.\ Statist.} {\bf 37}, 1211--1233 (1966)

\bibitem{KLPP97}
T.~Kurtz, R.~Lyons, R.~Pemantle, Y.~Peres:
\newblock A conceptual proof of the {K}esten-{S}tigum theorem for multi-type
  branching processes.
\newblock In K.~B. Athreya, P.~Jagers, eds, {\em Classical and Modern
  Branching Processes}, pp. 181--185, Springer, New York, 1997

\bibitem{Lind92}
T.~Lindvall:
\newblock {Lectures on the Coupling Method}.
\newblock Wiley, New York, 1992

\bibitem{LPP95}
R.~Lyons, R.~Pemantle, Y.~Peres:
\newblock Conceptual proofs of {L}log{L} criteria for mean behaviour of
  branching processes.
\newblock {\em Ann.~Prob.} {\bf 23}, 1125--1138 (1995)

\bibitem{Mitr70}
D.~Mitrinovic:
\newblock {Analytic inequalities}.
\newblock Springer, Berlin, 1970

\bibitem{Rock70}
R.~T. Rockafellar:
\newblock {Convex Analysis}.
\newblock Princeton University Press, Princeton, NJ, 1970

\bibitem{Stan04}
W.~Stannat:
\newblock On the convergence of genetic algorithms -- a variational approach.
\newblock {\em Probab.\ Theory Relat.\ Fields} {\bf 129}, 113--132 (2004)

\end{thebibliography}

\newcommand{\noopsort}[1]{} \newcommand{\printfirst}[2]{#1}
  \newcommand{\singleletter}[1]{#1} \newcommand{\switchargs}[2]{#2#1}

\end{document}